\documentclass[11pt,a4paper]{article}
\pdfoutput=1
\usepackage{jheppub} 
\usepackage{graphics,graphicx}
\usepackage{amsfonts,amsmath,amsthm,amssymb}
\makeatletter

\newcommand{\Rmnum}[1]{\expandafter\@slowromancap\romannumeral #1@}
\makeatother

\title{Unitary S Matrices With Long-Range Correlations and the Quantum Black Hole}

\author{Ratindranath Akhoury}

\affiliation{Michigan Center for Theoretical Physics, Randall Laboratory of Physics, University of Michigan, Ann Arbor, MI 48109-1120, USA}

\emailAdd{akhoury@umich.edu}

\abstract{We propose an S matrix approach to the quantum black hole in which causality, unitarity and their interrelation play a prominent role. Assuming the 't Hooft S matrix ansatz for a gravitating region surrounded by an asymptotically flat space-time we find a non-local transformation which changes the standard causality requirement but is a symmetry of  the unitarity condition of the S matrix. This new S matrix then implies correlations between the in and out states of the theory with the involvement of a third entity which in the case of a quantum black hole, we argue is the horizon S matrix. Effects of spacetime curvature and horizon are in fact introduced by this procedure which is seen to be a generalization of the Bogoliubov transformation. The analysis is performed within the Bogoliubov S matrix framework by considering a spacetime consisting of causal complements with a boundary in between. No particular metric or lagrangian dynamics need be invoked even to obtain an evolution equation for the full S matrix. Hawking's results are reproduced by restricting to low energy incoming modes at the horizon and the generalized hamiltonian of the horizon S matrix in this case is shown to be the generator of the Bogoliubov transformation. The modification of Bogoliubov causality at intermediate stages of black hole evaporation allows for a temporary violation of quantum mechanical no cloning theorems. In this way we find that the tension between information preservation and complementarity may be resolved provided the full quantum gravity theory either through symmetries or fine tuning forbids the occurrence of closed time like curves of information flow. Then, even if causality is violated near the horizon at any intermediate stage, a standard causal ordering may be preserved for the observer outside the black hole. The usefulness of our  formulation is that it appears well suited to understand unitarity at any intermediate stage of black hole evaporation. Moreover,  it is applicable generally to all theories with long range correlations including the final state projection models. As a nontrivial check, we use it in the perturbative context to analyze infrared divergences in QED and thereby reproduce the Faddeev-Kulish theory of asymptotic dynamics.}

\arxivnumber{}

\begin{document}

\maketitle

%%%%%%%%%%%%%%%%%%%%%%%%%%%%%%%%%%%%%%%%%%%%%%%%%%%%%%%%%%%%%%%%%%%%%%%%%%%%%%%%%%%
\section{Introduction}
The standard S matrix framework has been successful in describing most of the physical systems we encounter. The Hilbert space is a Fock space and the S matrix is a product of the in and out S matrices, i.e., $S=S_{\text {out}}S_{\text {in}}$. The validity of this standard framework is questioned by Hawking's discovery \cite{Hawking,Hawking2} of the failure of unitarity in black hole evaporation. In a series of papers, starting almost two decades ago, 't Hooft \cite{thooft1,stephens,thooft1.5,thooft2,thooft2.5} was the first to propose and to develop in a particular approximation scheme, a S matrix ansatz for the quantum black hole. Many concepts that we consider important today in black hole physics, like holography \cite{thooft-h,susskind-h} and entangled Hilbert spaces arose as an outcome of this program. Drawing intuition from the eikonal approximation where the transverse components of the momenta of the particles are neglected, he proposed a factorization of the black hole S matrix into three parts: $S_{\text{out}}S_{\text{horizon}}S_{\text{in}}$. The horizon S matrix was then constructed in the eikonal approximation including only the gravitational interactions and the correlation between the in and out states was explicitly verified. In these papers the correlations between the in and out states which are supposed to unitarize the black hole S matrix were attributed to large gravitational effects at the horizon.  Another ingredient central to the S matrix program and one which was simultaneously put forward by 't Hooft and Susskind, Thorlacius and Uglum is the idea of black hole complementarity \cite{thooft2,thooft2.5,thooft-c,susskind-c,Verlinde2}. 

In this paper we will adopt the viewpoint that black hole evaporation is different from, say, the decay of an elementary particle, or the ubiquitous example of the burning of paper, in the sense that long range correlations between in and out states arising from gravitational and other interactions are needed to restore unitarity.\footnote{This view is opposite to the philosophy inherent to the Fuzzball proposal, see for example \cite{Mathur}}  We develop a S matrix formulation which takes these correlations into account and is essential for understanding unitarity (the information loss problem) at any intermediate stage of black hole evaporation. Causality plays an important role in this analysis and we show that to get the desired correlations, there must necessarily be a modification of standard causality at the horizon for the time dependent cases. This modification at intermediate stages of black hole evaporation in turn allows for the a temporary evasion of quantum mechanical theorems requiring the monogamy of entanglement and no cloning. In this way we propose to reconcile the apparent tension \cite{Firewalls,Firewalls1} \footnote{For some earlier work see \cite{Braunstein,mathur2,Giddings1}} between information preservation and the principle of complementarity. In the process we have developed an S matrix theory of gravity in which effects of spacetime curvature and the horizon are introduced by a generalization of the Bogoliubov transformation and particle production through a generalization of thermo field dynamics\cite{thermofield,Israel}. The departures from standard causality are similar to those in final state projection models, \cite{maldacena, preskill1,preskill2} which are themselves easily accommodated in our formulation. The rest of this paper attempts to provide a platform for this summary.

The above considerations make it clear that our main task is the following: starting from the standard formulation we need to introduce correlations between the in and out states at the same time maintaining the unitarity of the S matrix. This can be done by introducing another unitary transformation which mediates between the in and the out states. For reasons which will soon become clear, this unitary will, in general, be called the boundary S matrix. This object is, in fact, a generalization of the Bogoliubov transformation and performs a dual role: It is the way that effects of spacetime curvature and horizon are introduced into what was the flat space S matrix, thereby generating the entanglement between the inside and outside of the black hole which drives the Hawking information loss argument. In addition it also is responsible for introducing correlations between the in and out states due to gravitational and other interactions. As we argue later, the boundary S matrix may be identified with just the horizon S matrix for the case of black holes. There are constraints from quantum information and from asymptotic boundary conditions on the nontriviality of this boundary S matrix. After all we do want to change the S matrix to take into account the entanglement across the horizon and the long range correlations, making sure at the same time that the unitarity condition is maintained. It is clear from quantum information theoretic  \cite{Buscemi,Nielsen} considerations that in order to introduce entanglement and long range correlations between the in and out states the boundary S matrix will be nonlocal in the sense that it will involve different regions of spacetime which are causal complements of each other. Another constraint comes from taking into account the conditions under which a S matrix can be changed by unitary transformations. From field  theoretical  considerations, in \cite{Chisholm} (and references therein)  it was argued that a unitary transformation does not change the S matrix if the transformations are local and if there are no boundary contributions. We will assume that this is true, in general, when both outer and inner boundaries are present even though in \cite{Chisholm} only the outer boundaries at large times were considered. We conclude then that the unitary transformation we seek in order to introduce correlations between the in and out states will be nonlocal and will be a boundary contribution. This naturally brings up the question: how will the introduction of these nonlocal unitary transformations with support on a boundary affect causality? Within an S matrix approach the relationship between unitarity and causality is to our knowledge best discussed using the Bogoliubov formulation \cite{Bogoliubov}. It is worthwhile to recall here that while ``unitarity" can be used both in the context of probability conservation and (no) information loss, it is only the latter which is an issue for the black hole problem. In the Bogoliubov approach to be discussed in this paper, the two are not unrelated on account of the fact that causality violation can give rise to information loss \cite{Deutsch,lloyd}.

 Most of the considerations of this paper are relevant for a spacetime divided into causal complements separated by a boundary which is the horizon. Broadly speaking, it develops an approach to study the unitary evolution of a S matrix with correlations between the in states and the out states with the involvement of a third entity, the boundary S matrix. In order to accomplish this, a disturbance propagating in the system is elevated to a space-time dependent ``coupling constant" ($g(x)$) and the evolution equation of the S matrix as a function of this variable is studied. In fact, our analysis, which is designed to incorporate the above mentioned correlations, is an adaptation of the Bogoliubov \cite{Bogoliubov} standard S matrix formalism and it is not necessary to specify a metric or any particular lagrangian dynamics. In addition to the S matrix ansatz, we also assume the validity of the standard causality requirement in asymptotically flat space-times with an energy momentum tensor satisfying the positive energy condition before any correlations are introduced through the boundary S matrix.  A transformation of a generalized Hamiltonian is found which changes the causality condition but is a symmetry of the unitarity condition of the S matrix. This transformation then connects the standard S matrix formulation to the one with correlations between the in and the out states. In this way the invariance of the S matrix unitarity condition is used to introduce the boundary S matrix which is responsible for the dynamics which correlates the in with the out states. A surprising and novel conclusion we arrive at is that the nonlocal nature of the boundary S matrix necessarily implies a violation of Bogoliubov causality near the horizon for the time dependent cases.  The interactions responsible could be gravitational in nature and are the same ones that are responsible for correlating  the in and out states. Our formalism also applies to other situations where long range correlations are important. Well known examples are provided in theories with massless particles: infrared divergences arise in perturbative gauge theories and gravity because interactions do not vanish at large times. The boundary S matrix approach developed in this paper reproduces the well known Faddeev-Kulish \cite{FK} characterization of the asymptotic S matrix in QED. This provides some confidence in our approach and also helps in understanding the meaning of the various terms in the evolution equation. In the Faddeev-Kulish theory the asymptotic states are not the Fock states but the coherent states which take into account the soft photon cloud surrounding a charged particle. The appearance of these coherent states which in fact form an over-complete basis is a novel feature with lessons for the black hole case. Some other important insights relevant for the black hole are pointed out. 

The implementation of this program allows us to draw many general conclusions about the dynamics of the horizon, about the nature of the in and out Hilbert spaces and about the implementation of holography and complementarity. The symmetry transformation of the unitarity condition that produces the new causality requirement is like a nonlocal gauge transformation with a generator of the form $e^{i\phi(g_1,g_2)}$, which depends on the disturbances ($g_1$ and $g_2$ respectively) in {\emph{both}} the in and the out regions (hence the nonlocality necessary for entanglement). Using the new causality condition, we also find a new evolution equation for the full S matrix in which $e^{i\phi}$ plays an important role. Special boundary conditions, which we call entangling boundary conditions must be imposed on $\phi$ which ensure that $e^{i\phi}$ is {\emph{not}} factorizable into a product of factors $(e^{i\phi_1(g_1)})_1(e^{i\phi_2(g_2)})_2$ in the two causally complementary regions ($1$ and $2$) respectively. Only in this case are nontrivial correlations introduced. The new causality condition may then be shown to lead to a similar factorization of the full S matrix that was proposed by 't Hooft \cite{thooft1,stephens,thooft1.5} ($S=S_{\text{out}}S_{\text{horizon}}S_{\text{in}}$) with $e^{i\phi}$ playing the role of the horizon S matrix. Perhaps the most intriguing conclusion of this paper is the realization that a violation of the Bogoliubov causality condition is forced upon us by requiring correlations between the in and the out states which define the S matrix. Causality violation was discussed in \cite{preskill1,preskill2} in the context of the solution of the black hole information puzzle proposed in \cite{maldacena}. There, it was argued that standard causal ordering may however be recovered under some reasonable conditions. Such a conclusion may be argued to hold in our  case too even though the mechanism for causality violation is seemingly different. This potentially places constraints on the full theory of quantum gravity which at intermediate stages may consequently allow for an evasion of the principles of the monogamy of entanglement and cloning. We will argue that in fact this connection with the work of \cite{maldacena} and \cite{preskill1,preskill2} is not coincidental. Until now we have been discussing how the symmetry transformation of the unitarity relation may be used to relate two theories, one with and the other without a horizon through the introduction of the horizon S matrix. However in a theory with a horizon already present, the same boundary S matrix can play the role of a unitarity transformation implementing the correlations needed to impose the final state boundary conditions in \cite{maldacena}. The formulation of this paper easily accommodates this situation and so it is not surprising that similar acausal information flow is an outcome of both cases. Causality also places strong restrictions on the applicability of a universal effective field theory description simultaneously for observers at the horizon and  for the outside observer. Using these ideas it is argued here that black hole complementarity is  realized without the introduction of firewalls \cite{Firewalls,Firewalls1} for the infalling observer provided the full quantum gravity theory forbids the occurrence of closed time like curves of information flow. Causality violation at the horizon is a consequence of other approaches seeking to resolve the black hole information paradox. In particular, see Ref.\cite{Giddings}, although the connection to the present work is unclear at the moment.

An evolution equation satisfied by the unitary S matrix is derived for both the in and out regions. As an application, the Hawking result is obtained in the semiclassical approximation, by limiting the incoming modes near the horizon to very low energy and ignoring quantum gravitational effects at the horizon. The restriction to low energy incoming modes at the horizon essentially makes the problem time independent and in this situation no causality violation is predicted. In fact, for reproducing the Hawking result, we find that the horizon S matrix just plays the role of generator of the time independent Bogoliubov transformation. The picture here is the same as in the original application of thermo field dynamics to the black hole\cite{Israel} with the two causally complementary regions of our approach playing the role of the doubled spaces in the thermo field approach. The full evolution equation allows us to go beyond this approximation and automatically shows how gravitational effects can introduce correlations between the in and out states with the potential of unitarizing the Hawking picture. Even though, the applications to black hole physics are the most interesting to the author, the formulation of S matrix theory developed in this paper is quite novel and is anticipated to have other applications in quantum information theory and statistical physics as well. 

The paper is organized as follows. In section~(\ref{secbogrev}) we review the flat space Bogoliubov approach to S matrix theory. Section~(\ref{Sgravity}) constitutes the essential part of the paper. We start with an overview of the procedure for developing the S matrix theory of gravity. Then in section~(\ref{Modcausal}) we show how a new  gauge structure exhibiting the symmetry of the unitarity relation can be used to introduce the boundary S matrix which however, changes the Bogoliubov causality condition. We then show how this leads to the factorization of the total S matrix  involving the boundary S matrix and the in and out S matrices. As emphasized many times, the introduction of the boundary S matrix can be thought of as a generalization of the Bogoliubov transformation which introduces the effects of space time curvature, and horizon into the flat space S matrix. Then in section~(\ref{evolution}) the evolution equation of the S matrix implied by the new causality condition is obtained. We conclude this section with a discussion of how the usual semiclassical results follow from the evolution equation.  In section~(\ref{Applications}) we explore the consequences of the new evolution equation of the S matrix. Hawking radiation with and without gravitational interactions included is discussed in section~(\ref{Hawking}). In section~(\ref{complementarity}) we develop an interpretation of the transformations which were used to introduce the boundary S matrix, and which are a symmetry of the unitarity condition. We also show how the proposal of \cite{maldacena} may be accommodated in our formulation. The consequences of Bogoliubov causality violation and the possibility of effective field theory descriptions are the subject of section~(\ref{causalityeft}) where we also explore the conditions for the realization of the black hole complementarity principle. The question of the commutation relations of operators with spacelike separation, one located inside and the other outside the black hole is also discussed. In section~(\ref{FKconnection}), the Faddeev-Kulish theory of asymptotic states for infrared divergences in perturbative gauge theories and gravity is seen to follow from the evolution equation. In this case also, the boundary S matrix acts as a ``dressing'' operator which dresses up the usual Fock space states to form coherent states representing a charged particle surrounded by a soft photon (or graviton) cloud.  We conclude with a summary of our results in section~(\ref{Conclusion}). In an appendix we solve the new evolution equation of the S matrix for a simplified situation and in this case make explicit the mechanism by means of which information is transferred from the in to the out region. The in and out states differ by a phase factor which involves an integral over the profile of the disturbance in the ``in" region. Accordingly, this phase stores information of the entire history.
%%%%%%%%%%%%%%%%%%%%%%%%%%%%%%%%%%%%%%%%%%%%%%%%%%%%%%%%%%%%%%%%%%%%%%%%%%%%%%%%%%%
\section{Review of Bogoliubov Causality }
\label{secbogrev}
The Bogoliubov formulation of causality within the S-matrix framework is a very convenient one specially for studies of its interrelation with the unitarity condition. Before we discuss the adaptation of this to include long range correlations, we will first review in flat space, the usual description of Bogoliubov causality \cite{Bogoliubov}. 

In their construction of the scattering matrix, Bogoliubov and Shirkov introduce an interaction region $G$ by making the coupling space time dependent, i.e., $g \rightarrow g(x)$. The function $g(x)$ takes values in the range $(0,1)$ and represents the extent of the switching on of the interaction. We can switch on an interaction or a disturbance of strength $g(x)$ by replacing the interaction lagrangian $L(x)$ by $g(x)L(x)$. Next divide the region $G$ into two complementary subregions $G_1$ and $G_2$. This division is such that all points in $G_1$ are in the past with respect to a certain time $t$ and all points in the subregion $G_2$ lie in the future with respect to $t$. Thus $g(x)$ can be represented as a sum of two functions: 
\begin{equation}
g(x) = g_1(x) + g_2(x),
\label{firstequation}
\end{equation}
with $g_1 \neq 0$ only for $x^0 < t$ and $g_2 \neq 0$ only for $x^0 > t$. The first step in determining the causality condition is to require that any event occurring in the system may exert influence on the evolution only in the future. Let us write the evolution of the S matrix, $S(g)$, in the form:
\begin{equation}
i{\delta S \over \delta g(x)} = i\left({\delta S \over \delta g(x)}S^{\dagger}\right)S.
\end{equation}
 It is convenient to introduce a generalized Hamiltonian operator by the following definition, 
\begin{equation}
H(x,g) = i{\delta S \over \delta g(x)}S^{\dagger}.
\label{hamop}
\end{equation}
The reason for this nomenclature \cite{Bogoliubov} will become clear later in this section (see also the discussion at the end of section (\ref{Modcausal})) when we show that in the semiclassical approximation, this object reduces to the usual Hamiltonian which dictates the time evolution of the S matrix in the interaction picture. The causality condition stated above can be mathematically expressed as,
\begin{equation}
{\delta \over \delta g_1(y)} \left({\delta S(g) \over \delta g_2(x)}S^{\dagger}(g)\right) = 0.
 \label{causality1}
\end{equation}
Using the space time points as the only labels, we may write this condition as,
\begin{equation}
{\delta \over \delta g(y)} \left({\delta S(g) \over \delta g(x)}S^{\dagger}(g)\right) = 0, x^0 > y^0.
\label{causalitys}
\end{equation}
We will refer to this as Bogoliubov causality. It is instructive to see how this condition can be expressed 
as the condition that the S-matrix factorizes in the following form: 
\begin{equation}
S(g) = S_2(g_2)S_1(g_1); G_2 > G_1.
\label{Sfactor}
\end{equation}
Indeed,  consider the case when the interaction in region $G_2$ is changed by an infinitesimal amount, then:
\begin{equation}
{\delta S(g) \over \delta g_2(x)} = {\delta S_2(g_2) \over \delta g_2(x)}S_1(g_1),
\end{equation}
where we have used the factorization property of the S-matrix which was alluded to earlier. Now we use the unitarity of the S- matrix, i.e., $SS^{\dagger} = S^{\dagger}S = 1$ to arrive at,
\begin{equation}
{\delta S(g) \over \delta g_2(x)}S^{\dagger}(g) = {\delta S_2(g_2) \over \delta g_2(x)}S_2^{\dagger}(g_2).
\end{equation}
The right hand side is independent of $g_1$, hence we again arrive at Eq.(\ref{causality1}) and 
Eq.(\ref{causalitys}). 
From Lorentz invariance, we can write down the covariant form of the causality condition:
\begin{equation}
{\delta \over \delta g(y)} \left({\delta S(g) \over \delta g(x)}S^{\dagger}(g)\right) = 0, x \gtrapprox y,
\label{Bcausality}
\end{equation}
where, the condition holds whenever $x$ is inside or on the future light cone of $y$ or when $x$ and $y$ are separated by a space like interval. We would like to emphasize that the causality condition and the factorization of the S-matrix Eq.~(\ref{Sfactor}) are closely linked. In order to make contact with the usual requirement of microcausality, we note that in terms of the generalized Hamiltonian the causality condition may be written as:
\begin{equation}
{\delta H(x,g) \over \delta g(y)} = i{\delta^2 S \over \delta g(y)\delta g(x)}S^{\dagger} + i H(x,g)H(y,g) = 0, x \gtrapprox y.
\end{equation}
From this it follows that, 
\begin{equation}
{\delta H(x,g) \over \delta g(y)} - {\delta H(y,g) \over \delta g(x)} = i [ H(x,g), H(y,g) ].
\label{microcausality}
\end{equation}
The usual condition of microcausality for the currents then follows from Lorentz invariance for points $x$ and $y$ that are space like separated:
\begin{equation}
 [ H(x,g), H(y,g) ] = 0 , x \approx y.
\end{equation}
It is worth pointing out that the Bogoliubov causality condition Eq.~(\ref{Bcausality}) is more general than the microcausality requirement since the former provides some information about the time-like region as well. We will see in the next section, that preservation of unitarity allows for a change of the causality condition, Eq.~(\ref{Bcausality}).

In concluding this short review, it will be relevant for future discussion to outline how the conditions of causality and unitarity of the S- matrix may be used to derive the usual Dyson expression for it in a lagrangian field theoretical formulation.For this purpose, One may expand the operator $S(g)$ in a power series in $g(x)$,
\begin{align}
S(g) = 1+&\sum_{n \geq 1} \frac{1}{n!}\int S_n(x_1,x_2,....,x_n)
g(x_1)g(x_2)....g(x_n)d^4x_1d^4x_2....d^4x_n,
\label{Sexpansion}
\end{align}
where, $S_n(x_1,x_2,....,x_n)$ is a polylocal operator.
In order to implement the causality condition it is very convenient to consider the generalized Hamiltonian introduced in Eq.~(\ref{hamop}). It has the expansion,
\begin{align}
H(x,g) = \sum_{n \geq 0} \frac{1}{n!}\int H_n(x,x_1,...,x_n)g(x_1)...g(x_n)d^4x_1...d^4x_n, \nonumber \\
H_n(x,x_1,...,x_n) = iS_{n+1}(x, x_1,....,x_n) + \nonumber \\
i\sum_{0\leq k\leq n-1} P\left({x_1,....,x_k} \over {x_{k+1},...,x_n}\right)
S_{k+1}(x, x_1,...x_k)S^{\dagger}(x_{k+1},...,x_n),
\end{align}
and $P$ is the complete symmetrization symbol and denotes the sum over all the ways of breaking up the $n$ points into sets of $k$ and $n-k$ points. The causality condition Eq.~(\ref{Bcausality}) then
implies that \cite{Bogoliubov} for $n\geq1$,
\begin{eqnarray}
H_n(x,x_1,...,x_n) = 0, \nonumber \\
\text{if for at least one}~ x_j (j = 1,...,n), ~x \geq x_j
\label{constraint1}
\end{eqnarray}

Using both the unitarity and the causality conditions, all the $S_n$ may be expressed in terms of $S_1$ alone \cite{Bogoliubov}. 

For example,
\begin{eqnarray}
S_2(x,y) = S_1(x)S_1(y),~x^0 > y^0  \nonumber \\
S_2(x,y) = S_1(y)S_1(x),~y^0 > x^0,
\end{eqnarray}
and so on. Then introducing a hermitian operator $\Lambda$, such that $S_1(x) = i\Lambda(x)$
the S-matrix may be written in the Dyson form:
\begin{equation}
S = T e^{\displaystyle i\int d^4x \Lambda(x)}.
\label{dyson}
\end{equation}
$\Lambda(x)$ is identified with the Lagrangian by considering the semiclassical correspondence\cite{Bogoliubov}. Finally with the above identification, we note for future reference that the expansion of the generalized Hamiltonian may be written as,
\begin{equation}
H(x,g) = H(x) + \sum_{n \geq 1} \frac{1}{n!}\int H_n(x,x_1,...,x_n)g(x_1)...g(x_n)d^4x_1...d^4x_n,
\end{equation}
where $H(x)$ is the usual Hamiltonian and the $H_n$ satisfy the constraint Eq.~(\ref{constraint1}).

%%%%%%%%%%%%%%%%%%%%%%%%%%%%%%%%%%%%%%%%%%%%%%%%%%%%%%%%%%%%%%%%%%%%%%%%%%%%%%%%%%%
\section{Extension to a S matrix Theory of Gravity}
\label{Sgravity}
In this section we will briefly explain our approach for extending the formalism of the previous section to develop a S matrix theory of gravity for asymptotically flat spacetimes enclosing a gravitating region. We will assume that in this case the S matrix can describe even the production and subsequent decay of a black hole and explore its consequences. This is the content of the 't Hooft S matrix ansatz \cite{stephens}. Even for the case of quantum field theory in a curved spacetime background, the differences with flat spacetime are quite significant. In particular, in the absence of Poincare invariance, the in and out vacuua are not the same. In this case, however, it is well known that the S matrix is the Bogoliubov transformation relating the two vacuua. Particle creation and the effects of the horizon can be included using the thermofield \cite{thermofield, Israel} formulation. An example is the Bogoliubov transformation that takes one from the vacuum of the parochial observers to the Hawking vacuum for the case of the eternal black hole. One may think of the Bogoliubov transformation as introducing the entanglement between the inside and the outside of a black hole which drives the Hawking information loss argument. Here it is worth mentioning that the thermofield formulation together with insights from the Tomita-Takesaki theory \cite{Ojima,Landsman,Raju1} can allow for a construction of the states inside the black hole for the equilibrium case. In the following subsection we will introduce a generalization of the Bogoliubov transformation which goes beyond the external field framework.

The effects of spacetime curvarture and horizon are introduced through a unitary operator which we call the boundary S matrix or the horizon S matrix. This boundary S matrix will play the role of the generalization of the Bogoliubov transformation and in the following will be written in a form (using the Magnus representation\cite{Magnus}) $\omega(g_1,g_2)= e^{i\phi(g_1,g_2)}$, with $\phi$  the generator of the transformation. For the black hole case, the boundary S matrix will play a dual role.  First, it introduces the entanglement between the inside and the outside of the black hole which drives the Hawking information loss argument. In addition it also implements the long range correlations due to gravitational and other interactions necessary for the unitarization of the total S matrix. It depends on the ``couplings" $g_1$ and $g_2$ in two regions which are causal complements of each other (see Eq.~(\ref{firstequation}) and the discussion following). As we will see in the next section, in order to act nontrivially, the boundary S matrix must satisfy the boundary conditions, $\omega(g_1,0)=\omega(0,g_2)=\omega(0,0)=0$. How can the boundary S matrix be introduced into the discussion of section~(\ref{secbogrev})? We want to change the S matrix to a new one which incorporates the above mentioned correlations, maintaining the unitarity of the new S matrix at the same time. We show in the next section how this is done using a gauge like symmetry of the unitarity condition. In this way, the unitarity of the new S matrix is preserved but the causality condition is changed. The violation of Bogoliubov causality is thus linked to the preservation of unitarity through long range correlations. The new causality condition, in turn implies an evolution equation for the new S matrix some of whose solutions we discuss in the paper.

More specifically we find that the factorization of old S matrix as shown in Eq.~(\ref{Sfactor}) is changed to:
\begin{equation}
S(g) \rightarrow S'(g) = \omega(g_1,g_2)S_2(g_2)S_1(g_1); G_2 > G_1.
\end{equation}
This may be put in the form of the  factorization of the S matrix for black hole formation and evaporation as proposed in \cite{thooft1,stephens,thooft1.5,thooft2,thooft2.5} by invoking some preferences about locality of the description. We discuss this also in the next section.Finally we show in section~(\ref{Hawking}) that when limited to low energy incoming modes at the horizon, our formulation reduces to the discussion of Ref.~\cite{Israel} with the regions labelled $1$ and $2$ being just the out and in regions of the Kruskal manifold. More significantly, $\phi$ turns out to be just the generator of the Bogoliubov transformation that takes one from the vacuum of the parochial observers to the Hawking vacuum. The usual picture within the thermo field formalism is thus recovered for this equilibrium situation with the operators in the regions labelled $1$ and $2$ playing the roles of the doubled thermo field ones, i.e., the tilde and the non-tilde operators\cite{thermofield,Israel}.

In concluding this overview, it is worth pointing out that we do not invoke any particular metric to define the S matrix for gravity or to obtain an evolution equation for it. All effects of spacetime curvature and the horizon appear as different types of correlations introduced through the boundary S matrix. In this sense one may say that we are presenting an explicit realization of how to obtain spacetime from entanglement \cite{Swingle,Maldacena2}. We comment further about this in section~(\ref{complementarity}) where we also speculate about a connection between quantum gravity and the continuous permutation group.

%%%%%%%%%%%%%%%%%%%%%%%%%%%%%%%%%%%%%%%%%%%%%%%%%%%%%%%%%%%%%%%%%%%%%%%%%%%%%%%%%%%

\subsection{A Modification of the Causality Relation}
\label{Modcausal}
While the Bogoliubov formalism of section~(\ref{secbogrev}) works very well for most of the problems encountered, there are instances like the Hawking \cite{Hawking} calculation of black hole radiance where problems with unitarity are encountered. 't Hooft \cite{thooft1,stephens,thooft1.5,thooft2,thooft2.5} has pioneered an S matrix approach for the quantum black hole where he shows using the case of eikonal scattering that the conventional approach needs to be modified. In particular, factorizations of the S matrix of the form Eq.~(\ref{Sfactor}) must be replaced by a more general one representing correlations between the in and out S matrices that is induced by the gravitational interactions of the horizon S matrix. There are several questions that arise. How can a third entity, the boundary S matrix, be introduced into the Bogoliubov formalism and how can the causality condition Eq.~(\ref{causalitys}) be changed so as to introduce the above mentioned correlations maintaining the unitarity of the S matrix at the same time? Is there a symmetry transformation that allows us to go from the case without the boundary S matrix to the case with it? Is this symmetry spontaneously broken, or more generally, in this new form of the S matrix can we say something about the ground state of the full theory? We will attempt to provide a positive answer to all these questions. We should also point out another reason why the Bogoliubov formalism is so well suited for this analysis. First, we make the simple observation that introducing a space time dependent measure $g(x)$ for a disturbance promotes the coupling to be a spurion. This spurion concept though not central in the previous section plays a role here. In particular, the evolution of the S matrix can be studied as a function of the coupling once introduced in this manner, without explicitly introducing the concept of time. As we show later, time emerges in the context of semiclassical reasoning. As is well known, these aspects are important when discussing the physics in theories involving gravitation. As discussed in the Introduction we will now consider an asymptotically flat spacetime which encloses a gravitating system and the energy momentum tensor satisfies the weak energy condition. As a notational convenience, for the rest of this paper, $[d^4x]$ will always refer to the invariant volume element.

We begin by exploring he relation between causality and unitarity of the S-matrix. We will demand that the S-matrix is unitary, and then show that we still have freedom to introduce the boundary S matrix refered to in the introduction thereby altering the analogue of the causality condition Eq.~(\ref{causalitys}). This alteration  however, introduces interesting correlations between the past and future light cones of the type which were first discussed by 't Hooft \cite{thooft1,stephens,thooft1.5,thooft2,thooft2.5} in the context of the S matrix approach for the quantum black hole. The altered causality condition implies a modification of the factorization Eq.~(\ref{Sfactor}) of the S- matrix.

Consider the analogue of the causality relation Eq.~(\ref{causalitys}). Combining the two regions $x > y$ ($x$ lies to the future of a spacelike surface through $y$)and $x < y$, we have:
\begin{align}
\theta(x,y){\delta \over \delta g(y)}\left({\delta S \over \delta g(x)}S^{\dagger}\right) + \theta(y,x){\delta \over \delta g(x)}\left({\delta S \over \delta g(y)}S^{\dagger}\right) = 0
\label{Causality1}
\end{align}
\begin{align}
{\delta^2 S \over \delta g(y)\delta g(x)}S^{\dagger} + \theta(x,y){\delta S \over \delta g(y)}{\delta S^{\dagger} \over \delta g(x)} + \theta(y,x){\delta S \over \delta g(y)}{\delta S^{\dagger} \over \delta g(x)} = 0.
\label{Causality2}
\end{align}
In the above $\theta(x,x')$ is the chronological step function defined such that $\theta(x,x')=1$ if $x$ lies to the future of a spacelike surface through $x'$ and is zero otherwise.
We would like to study the hermitian and anti-hermitian parts of this equation separately. For this purpose, we note that the hermitian conjugate of Eq.~(\ref{Causality2}) is:
\begin{align}
S{\delta^2 S^{\dagger} \over \delta g(y)\delta g(x)} + \theta(x,y){\delta S \over \delta g(x)}{\delta S^{\dagger} \over \delta g(y)} + \theta(y,x){\delta S \over \delta g(y)}{\delta S^{\dagger} \over \delta g(x)} = 0.
\label{Causality2hc}
\end{align}
Clearly the hermitian part of Eq.~(\ref{Causality2}) is simply,
\begin{align}
{\delta^2 (SS^{\dagger}) \over \delta g(y)\delta g(x)} = 0.
\label{hermitian}
\end{align}
Thus Bogoliubov causality already contains unitarity with suitable boundary conditions.
For comparative purposes, of relevance later on, we also write down the anti-hermitian part of Eq.~(\ref{Causality2}):
\begin{align}
{\delta^2 S \over \delta g(y)\delta g(x)}S^{\dagger} - S{\delta^2 S^{\dagger} \over \delta g(y)\delta g(x)}  \nonumber \\
- (\theta(x,y)-\theta(y,x))[{\delta S \over \delta g(x)}{\delta S^{\dagger} \over \delta g(y)} - {\delta S \over \delta g(y)}{\delta S^{\dagger} \over \delta g(x)}] = 0
\label{antihermitian}
\end{align}
If we seek a modification of the Bogoliubov causality condition keeping the S- matrix strictly unitary, then clearly the hermitian part of Eq.~(\ref{Causality2}), i.e., Eq.~(\ref{hermitian}) must be left unaltered. The only freedom we have is the possibility of nontrivially changing the anti-hermitian part. We next demonstrate how this can be done.

Clearly, The hermitian part of this equation is unaltered if to the right hand side of Eq.~(\ref{Causality2})
we add $i\chi(g_1, g_2)$, where $\chi$ is hermitian, i.e., $\chi^{\dagger} = \chi$ and we have reverted back to the notation of Eq.~(\ref{causality1}). Note that we look for a function that depends both on $g_1$ and $g_2$ because we would like in this way to introduce a correlation between the ``in" S matrix $S_1(g_1)$ and the ``out" S matrix $S_2(g_2)$ that appear in Eq.~(\ref{Sfactor}). In general, the two couplings $g_1$ and $g_2$ will be non-zero respectively in the two regions which are causal complements of each other. In order to see what this change means we will first consider the simpler case when $\chi$ commutes with the S-matrix. We will call this the abelian case. The results obtained in this simpler case will illustrate the kind of situations we encounter which are not covered by the standard analysis of Section~(\ref{secbogrev}). Following that we will take up the most general case.
%%%%%%%%%%%%%%%%%%%%%%%%%%%%%%%%%%%%%%%%%%%%%%%%%%%%%%%%%%%%%%%%%%%%%%%%%%%%%%%%%%%
%\subsection{Modified Causality Condition: The Abelian Case}

The fact that unitarity is unaffected if to the right hand side of Eq.~(\ref{Causality2}) we may add $i\chi$ can be much better expressed as an invariance of the unitarity relation, i.e., If $S \rightarrow S'$ in the manner shown below then both $SS^{\dagger}=S^{\dagger}S=1$ and 
$S'S'^{\dagger}=S'^{\dagger}S'=1$. The transformation relating $S$ and $S'$ is:
\begin{equation}
 {\delta S' \over \delta g(x)}S'^{\dagger} =  {\delta S \over \delta g(x)}S^{\dagger} -
e^{i\phi}{\delta \over \delta g(x)}e^{-i\phi},
\label{ginvariance}
\end{equation}
where, $\phi$ is a functional of the couplings $g_1$ and $g_2$. By this we mean that if we express ${\delta S \over \delta g(x)}S^{\dagger}$ as defined by Eq.~(\ref{ginvariance}) in terms of ${\delta S' \over \delta g(x)}S'^{\dagger}$ in Eq.~(\ref{Causality1}), then the unitarity relation Eq.~(\ref{hermitian}) is still obtained with the new S matrix $S'(g)$ replacing $S(g)$.
Eq.~(\ref{ginvariance}) has the appearance of an abelian gauge transformation for the generalized hamiltonian discussed in the previous section, $H(x,g) = i{\delta S \over \delta g(x)}S^{\dagger}$ with the coupling constant playing a role similar to the space-time variables. A crucial difference is that here $\phi(g_1,g_2)$ is nonlocal for reasons discussed in the introduction, namely that this is necessary for introducing entanglement.
From Eq.~(\ref{ginvariance})  and Eq.~(\ref{Causality1}) we see that even though unitarity of the S matrix is preserved, the generalized Bogoliubov causality condition is changed to (compare with Eq.~(\ref{causality1}) with $G_2>G_1$) :
\begin{equation}
 {\delta \over \delta g_1(y)}\left({\delta S(g) \over \delta g_2(x)}S^{\dagger}(g) - i{\delta \phi  \over \delta g_2(x)}\right) = 0.
 \label{newcausality}
\end{equation}
Henceforth, for notational simplicity,  if there is no ambiguity, we will drop the prime on $S$. 
This equation will clearly violate the factorization condition Eq.~(\ref{Sfactor}). 
In fact, we can show that the change the factorization condition is:
\begin{align}
S(g_1+g_2) = \omega(g_1,g_2)S_2(g_2)S_1(g_1);~G_2>G_1,
\label{omegafactor}
\end{align}
with a unitary $\omega$.
Then, as before, considering an infinitesimal change in the interaction in region $G_2$,
\begin{equation}
{\delta S(g) \over \delta g_2(x)} = \omega(g_1,g_2){\delta S_2(g_2) \over \delta g_2(x)}S_1(g_1) + {\delta \omega \over \delta g_2(x)}S_2(g_2)S_1(g_1).
\end{equation}
From this it follows that,
\begin{equation}
{\delta S(g) \over \delta g_2(x)}S^{\dagger}(g) = {\delta S_2(g_2) \over \delta g_2(x)}S_2^{\dagger}(g_2) + {\delta \omega \over \delta g_2(x)}\omega^{\dagger}.
\end{equation}
 We can write this in the form of the causality requirement, Eq.~(\ref{newcausality})
 \begin{equation}
{\delta \over \delta g_1(y)}\left({\delta S(g) \over \delta g_2(x)}S^{\dagger}(g) - \left({\delta \omega  \over \delta g_2(x)}\right)\omega^{\dagger}\right) = 0,
\label{cause}
\end{equation}
thus establishing the equivalence of Eq.~(\ref{newcausality}) and Eq.~(\ref{omegafactor}).
We conclude that the condition that replaces Eq.~(\ref{Sfactor}) is, for the same region, 
(by identifying $\omega(g_1,g_2) = e^{i\phi(g_1,g_2)}$)
\begin{equation}
S(g_1+g_2) = e^{i\phi(g_1,g_2)}S(g_2)S(g_1).
\label{newSfactor}
\end{equation}
Note that the S-matrix is no longer factorizable in a separable form. The quantity $e^{i\phi(g_1,g_2)}$ enforces the entanglement between the future and the past regions and is the boundary S matrix referred to earlier. As the entangling transformation must be nonlocal, it is clear that $\phi(g_1,g_2)$ must satisfy the following boundary conditions which will also be assumed for the general case:
\begin{equation}
\phi(g,0) = \phi(0,g) = 0~;~ \phi(0,0) = 0.
\label{bc}
\end{equation}
Note also that a simple factorization $\omega(g_1,g_2) = \omega_1(g_1)\omega_2(g_2)$ will not be sufficient. Apart from not satisfying the boundary conditions, for this ansatz the left hand side of Eq.~(\ref{cause}) will then trivially reduce to the old causality condition Eq.~(\ref{newcausality}):
\begin{equation}
 {\delta \over \delta g_1(y)}\left(\left({\delta \omega  \over \delta g_2(x)}\right)\omega^{\dagger}\right) 
=0={\delta \over \delta g_1(y)}\left({\delta S(g) \over \delta g_2(x)}S^{\dagger}(g)\right).
\end{equation} 
Thus the boundary conditions Eq.~(\ref{bc}) are essential for the program outlined here to work.
This will be true also in the general case which will be discussed soon and appears to be a general feature of the introduction of entangling unitary transformations.
Thus we see that a modification of Bogoliubov causality such that the S- matrix remains unitary, necessarily will violate the factorization property Eq.~(\ref{Sfactor}) of the S-matrix. This is a consequence of introducing correlations between the two causally complementary regions through the introduction of  the new entity $\phi(g_1,g_2)$. The nonseparable factorization of the S-matrix of the kind in Eq.~(\ref{newSfactor}) was first proposed by 't Hooft \cite{thooft1, stephens,thooft1.5,thooft2,thooft2.5} in the context of eikonal scattering at large center of mass energies and small momentum transfer within a field theoretic framework of gravity. There, $\omega$ was identified with the horizon S-matrix. We should emphasize that Eq.~(\ref{Sfactor}) differs from the 't Hooft factorization in the placement of the factor $e^{i\phi}$. We will discuss this further in the analysis of the general case below.

A glance at Eq.~(\ref{newSfactor}) suggests that the invariance of the unitarity relation, Eq.~(\ref{ginvariance}), has representations in the abelian case which are similar to the projective representations of groups. As we will see soon, this invariance has a ``non-abelain" extension as well. In appendix A, we will use the similarity with the projective representations to show that the net effect of the change Eq.~(\ref{newSfactor}) for the abelian case will be to shift the out states, for example, by a coupling constant dependent phase factor. This phase factor stores the entire history since it involves an integral over the disturbance in the in region. Thus, the mechanism by means of which information is transferred to the out states is indicated for this simple situation.

Next we will remove the abelian restriction imposed earlier and for this general case we will clearly outline the situations when this new form of the S matrix must be used. We would like to emphasize that
here as well as for the more general case, Eq.~(\ref{newSfactor}) arises in the context of a general S-matrix framework without reference to any specific field theory or string theory and neither is there any restriction to a perturbative analysis.

The above analysis provides us with some insight into what kind of physical situations the generalized causality conditions can describe. In Section~(\ref{secbogrev}), we reviewed the standard formulation of the S matrix in which microcausality was straightforwardly satisfied and the regions $G_1$ and $G_2$ have no special correlations with each other. The space of states are the usual Fock states. While this is what is normally encountered in quantum field theory, there are situations where this description is inadequate. We would like to have a unitary S matrix description of two regions $G_1$ and $G_2$ which are causal complements of each other and are separated by a boundary which is the horizon. We would like to include the possibility of  correlations in the quantum theory between these regions through the inclusion of a third entity $\phi(g_1,g_2)$ associated with a boundary region, where $g_1$ and $g_2$ are the nonvanishing couplings in the two regions as before. As we will see the nature of the Hilbert spaces in the generalization of Eq.~(\ref{newSfactor}) is much more complicated in that the above mentioned regions are highly entangled and therefore not independent. This feature will be essential for the unitarity of the S matrix. Fortunately for our subsequent analysis, we will not have to specify exactly the nature of these Hilbert spaces. Whatever assumptions are needed will be discussed when the generalization of Eq.~(\ref{newSfactor}) is obtained below. 

\begin{figure}
\begin{center}
\includegraphics[width=3.5in]{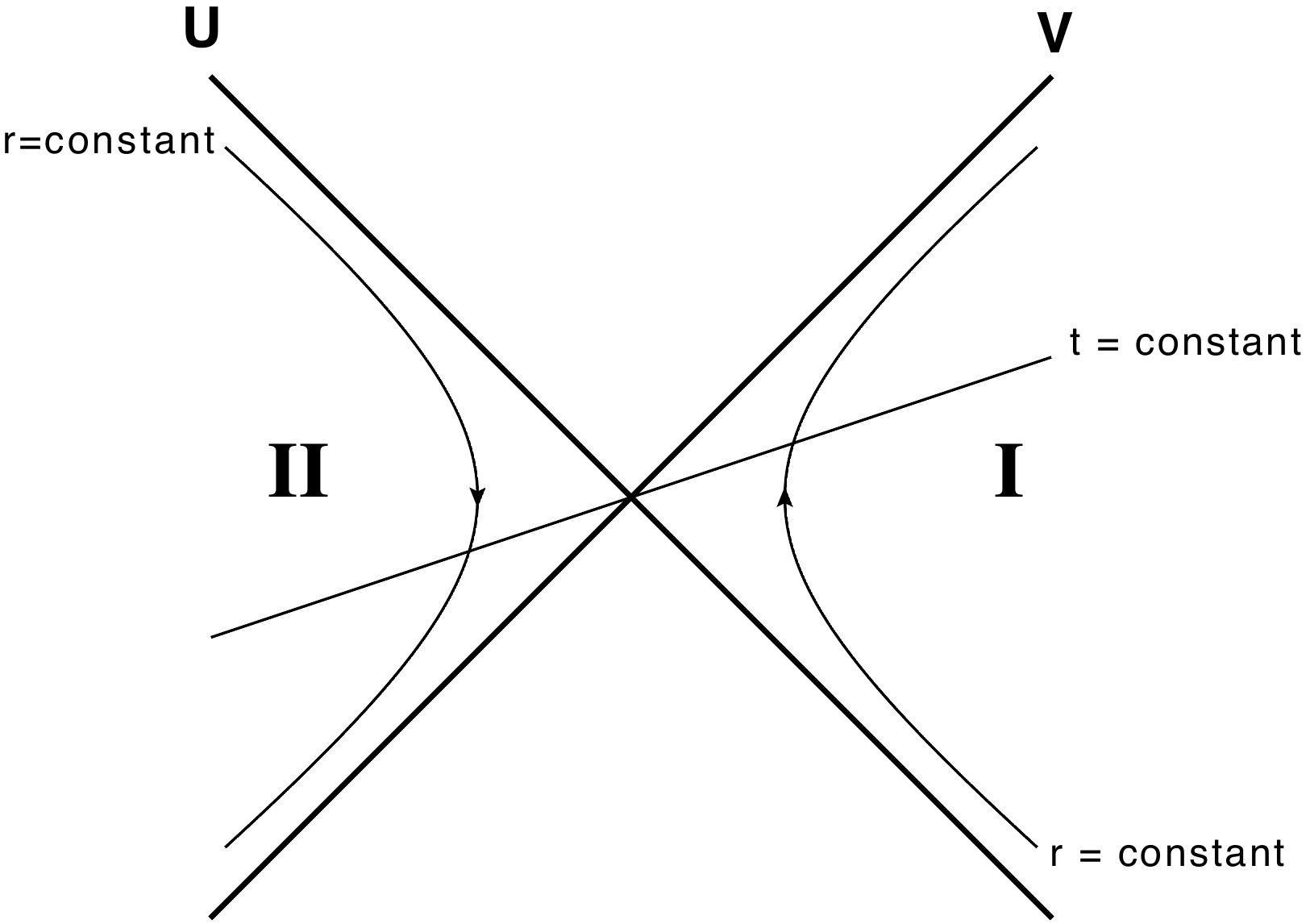}
\caption{Kruskal Manifold}
\end{center}
\label{ForAkhoury}
\end{figure}

In order to get a picture of what is being talked about it is worthwhile to give a concrete example of such a region which will be discussed in some more detail in a leading semiclassical approximation and beyond in section~(\ref{Hawking}). For the example we have in mind, this description is good at late times after the formation of a big black hole (a black hole which has settled down to a stationary state or an eternal black hole). The regions $G_2$ and $G_1$ could respectively be the standard regions \Rmnum{1} (including region \Rmnum{3}) and \Rmnum{2} (including region \Rmnum{4}) in the Kruskal manifold (see Fig.~(\ref{ForAkhoury})) or in the near horizon limit of a massive black hole,  they could represent the right and left Rindler wedges. The boundaries in either case are the future and past event horizons. We will call the region \Rmnum{1} as the out region and \Rmnum{2} as the in region. Thus we use the black hole metric for the in region and the white hole one for the out region. The spacetime dependent couplings  which are non-vanishing in the respective regions will be labeled $g_{\Rmnum{1}}$ and $g_{\Rmnum{2}}$ and $\phi(g_{\Rmnum{1}}, g_{\Rmnum{2}})$ would be associated with the dynamics of the inward and outward moving particles close to the horizon.  Another related physical situation we could describe using the formulation of this paper is the Unruh effect \cite{unruh,Wald} where we would use the Rindler coordinatization of Minkowski space with the left and right Rindler wedges being the in and out regions. The original calculation of the Unruh effect was for an eternally accelerating observer and the same time independent framework as discussed for the stationary black hole is applicable. However, as we will see soon, this analysis can be extended to the case of non-eternal observers as well.
 
 Though black hole physics is the most natural situation suited for the analysis of this paper, we will also consider another possibility where the boundary regions are surfaces at times $\pm \infty$ in Minkowski space. $\phi$ has support only on these surfaces. In this situation which is relevant for infrared dynamics  of say perturbative gauge theories and quantum gravity, the infrared degrees of freedom live on these surfaces and the description of $\phi$ is non-local. In section~(\ref{FKconnection})we will show how to recover the Faddeev-Kulish \cite{FK} theory of asymptotic states for infrared divergences using our formalism. Again a Fock space description is not appropriate and the relevant asymptotic states here are the coherent states representing charged particles surrounded by a soft photon cloud. Unless otherwise stated, we will be referring to the black hole case henceforth and return to the Faddeev-Kulish theory in section~(\ref{FKconnection}).

We are now ready to discuss the most general invariance of the unitarity relation, $SS^{\dagger}=S^{\dagger}S=1$. We will find that the results of the previous abelian analysis carry over with appropriate generalizations. Referring to Eq.~(\ref{ginvariance}), we would like to remove the restriction that $\phi$ commutes with the other elements of the S-matrix and further we consider the possiblity of multi-component dynamical variables in the sense that $\phi$ could be matrix valued. In this general case we would also like to write the boundary S matrix  in a form where it appears to be manifestly unitary. Suppose that $\omega(g_1,g_2)$ is such an operator, then it satisfies an evolution equation of the form, 
\begin{align}
i{\delta \omega \over \delta g(x)} = H_h(x,g)\omega \nonumber \\
H_h(x,g) = i{\delta \omega \over \delta g(x)}\omega^{\dagger}.
\label{horizon-H}
\end{align}
We choose to write $\omega$ in the Magnus representation \cite{Magnus}, i.e., 
\begin{align}
\omega(g_1,g_2) = e^{i\phi(g_1,g_2)},
\end{align}
where $\phi$ is hermitian and involves the integral of $H_h$ and its multiple commutators. We also use the entangling boundary condition Eq.(\ref{bc}).

It is straightforward to see that Eq.~(\ref{ginvariance}), can be generalized to,
\begin{equation}
{\delta S' \over \delta g(x)}S'^{\dagger} = e^{i\phi}{\left({\delta S \over \delta g(x)}S^{\dagger}\right)}e^{-i\phi} - e^{i\phi}{\delta \over \delta g(x)}e^{-i\phi}.
\label{genginvariance}
\end{equation}
To check that this leaves the unitarity relation unaltered, we note that the above transformation implies,
\begin{equation}
{\delta S \over \delta g(x)}S^{\dagger} = e^{-i\phi}{\left({\delta S' \over \delta g(x)}S'^{\dagger}\right)}e^{i\phi} - e^{-i\phi}{\delta \over \delta g(x)}e^{i\phi}.
\label{inverse}
\end{equation}
Clearly the causality condition is changed but we will check that unitarity is preserved.
Next we substitute for 
 ${\delta S \over \delta g(x)}S^{\dagger}$ in Eq.~(\ref{Causality1}) the right hand side of Eq.~(\ref{inverse}) (and a similar expression with $x$ and $y$ interchanged), add the hermitian conjugate of the resulting expression and verify that this implies $S'S'^{\dagger}=S'^{\dagger}S'=1$. Some useful intermediate results are:
\begin{align}
{\delta \over \delta g(y)}\left({\delta S \over \delta g(x)}S^{\dagger}\right) =  
e^{-i\phi} \big(
 {\delta \over \delta g(y)}\left({\delta S' \over \delta g(x)}S'^{\dagger}\right)  \nonumber \\
 + \left[ {\delta S' \over \delta g(x)}S'^{\dagger} , \left({\delta \over \delta g(y)}e^{i\phi}\right)e^{-i\phi} \right] 
- {\delta \over \delta g(y)}\left(\left({\delta \over \delta g(x)}e^{i\phi}\right)e^{-i\phi} \right)   
\nonumber \\
- \left[ \left({\delta \over \delta g(x)}e^{i\phi}\right)e^{-i\phi} ,  \left({\delta \over \delta g(y)}e^{i\phi}\right)e^{-i\phi} \right] \big) e^{i\phi}.
  \label{j-prime}
\end{align}
From this we can form the quantity,
\begin{align}
\theta(x,y)\left({\delta \over \delta g(y)}\left({\delta S \over \delta g(x)}S^{\dagger}\right)\right) \nonumber \\ 
+ \theta(y,x)\left({\delta \over \delta g(x)}\left({\delta S \over \delta g(y)}S^{\dagger}\right)\right) = 0.
\label{causality-prime}
\end{align}
Upon adding to Eq.~(\ref{causality-prime}) its hermitian conjugate we get,
\begin{align}
{\delta^2 \over \delta g(x) \delta g(y)}\left(S'S'^{\dagger}\right) + \nonumber \\
\theta(x,y)\left[{\delta \over \delta g(x)}\left(S'S'^{\dagger}\right) , \left({\delta \over \delta g(y)}e^{i\phi}\right)e^{-i\phi}\right] + \nonumber \\
\theta(y,x)\left[{\delta \over \delta g(y)}\left(S'S'^{\dagger}\right) , \left({\delta \over \delta g(x)}e^{i\phi}\right)e^{-i\phi}\right] = 0,
\end{align}
clearly indicating the preservation of unitarity of the S-matrix. The causality condition, however, is modified from the abelian case. 
To obtain the form of the causality relation in a suggestive form, let us introduce a derivative which measures the total change in a functional $F$ of the coupling constant $g$ as the latter itself evolves:
\begin{equation}
{DF(g) \over Dg(x)} = {\delta F \over \delta g(x)} + i[i \left({\delta e^{i\phi} \over \delta g(x)}\right)e^{-i\phi}, F]
\label{covariant-derivative}
\end{equation}
In fact, from Eqs.~(\ref{j-prime}) and (\ref{causality-prime}) and using Eq.~(\ref{covariant-derivative}) we can write the generalized causality condition for the case $x > y$ as, (compare with  Eq.~(\ref{cause}))
\begin{equation}
{D \over Dg(y)}\left[{\delta S' \over \delta g(x)}S'^{\dagger} - \left({\delta \over \delta g(x)}e^{i\phi}\right)e^{-i\phi}\right] = 0.
\label{geometriccausality}
\end{equation}
This is the modified causality condition and together with the corresponding one for $y > x$ describes the new dynamics of the S-matrix.

 Let us now see what this implies regarding the factorization of the S-matrix. For this it is more convenient to revert back to the form of the new causality condition Eq.~(\ref{causality-prime}) for $x>y$ i.e., : 
\begin{equation}
{\delta \over \delta g(y)}\left[e^{-i\phi}\left({\delta S \over \delta g(x)}S^{\dagger} - \left({\delta \over \delta g(x)}e^{i\phi}\right)e^{-i\phi}\right)e^{i\phi}\right] = 0.
\label{newgendynamics2}
\end{equation}
{\emph{Again, for notational convenience in the above and henceforth, we will drop the primes in $S$ whenever it is unambiguous}}.
This is the differential form, and we would like to obtain the integral version. Consider the ansatz of the previous section except that now $\omega$ in Eq.~(\ref{omegafactor}) is an operator which does not commute with the S-matrix:
\begin{equation}
S(g_1+g_2) = \omega(g_1,g_2)S_2(g_2)S_1(g_1);~G_2>G_1.
\end{equation}
Differentiating with respect to $g_2(x)$ as before we get,
\begin{equation}
{\delta S(g) \over \delta g_2(x)} = \omega(g_1,g_2){\delta S_2(g_2) \over \delta g_2(x)}S_1(g_1) + {\delta \omega \over \delta g_2(x)}S_2(g_2)S_1(g_1),
\end{equation}
 for a unitary $\omega$. Multiplying the right hand side of this by $S^{\dagger}=S_1^{\dagger}(g_1)S_2^{\dagger}(g_2) \omega^{\dagger}(g_1,g_2)$ gives,
\begin{equation}
{\delta S(g) \over \delta g_2(x)}S^{\dagger}(g) = {\delta \omega \over \delta g_2(x)}\omega^{\dagger} + \omega(g_1,g_2){\delta S_2(g_2) \over \delta g_2(x)}S_2^{\dagger}(g_2)\omega^{\dagger}.
\end{equation}
It immediately follows that 
\begin{equation}
\omega^{\dagger}\left[{\delta S(g) \over \delta g_2(x)}S^{\dagger}(g) - {\delta \omega \over \delta g_2(x)}\omega^{\dagger}\right]\omega= {\delta S_2(g_2) \over \delta g_2(x)}S_2^{\dagger}(g_2),
\end{equation}
and therefore that,
\begin{equation}
{\delta \over \delta g_1(y)}\left(\omega^{\dagger}\left[{\delta S(g) \over \delta g_2(x)}S^{\dagger}(g) - {\delta \omega \over \delta g_2(x)}\omega^{\dagger}\right]\omega\right)= 0.
\end{equation}
This is just Eq.~(\ref{newgendynamics2}) with the replacement $\omega(g_1,g_2) =
e^{i\phi(g_1,g_2)}$.
Thus the condition that replaces Eq.~(\ref{Sfactor}) is, for the same region, 
\begin{equation}
S(g_1+g_2) = e^{i\phi(g_1,g_2)}S_2(g_2)S_1(g_1),
\label{newgenSfactor}
\end{equation}
which is similar in form to the abelian case except that $\phi$ can be both an operator that does not commute with the S-matrix and it can be matrix valued. At this point we would again like to remind the reader that a factorization of the entangling unitary in the form: $\omega(g_1,g_2)=\omega_1(g_1)\omega_2(g_2)$ is not sufficient to get entanglement. As discussed after Eq.~(\ref{bc}), this factorization will give the trivial solution which is essentially the standard causality relation without the correlations. {\emph{Thus, it is important to realize that requiring nontrivial long range correlations necessarily implies a change in the causality condition}}. Another point worth emphasizing is that since $\phi$ always arises in the form $\delta e^{i\phi} \over \delta g(x)$, the causality violation is present only in cases where the dependance of $\phi$ on $g(x)$ is dynamical and evolving in time. 

Another point worth emphasizing here is that in deriving Eq.~(\ref{newgenSfactor}) we have assumed that $\omega(g_1,g_2)$, $S_1$ and $S_2$ are separately unitary. There may be obstructions to this apart from purely kinematic ones. For example, it could happen that a hermitian $\phi(g_1,g_2)$ cannot be constructed because of the presence of singularities. A singularity, for instance may be produced if the causality violation is due to the presence of closed time-like curves of information flow. For an example of this in the context of AdS/CFT see \cite{caldarelli,Milanese}. These closed time-like curves would cause information loss \cite{Deutsch,lloyd}. Thus, the full theory of quantum gravity must forbid such situations in order to have a unitary S matrix description. We will return to this point in section (\ref{causalityeft}).

Before proceeding, we would like to explain further our reasons for identifying $e^{i\phi}$ with the horizon S matrix. This involves our recipe for where the boundary  S matrix should be located in this picture in order to be identified with the horizon S matrix, i.e., we would like to discuss the placement of the entangling unitary $e^{i\phi}$. An important ingredient is our preference for preserving locality as much as possible. We have seen that to produce a unitary S matrix we have two possibilities. The conventional picture, and one relevant for black holes  in which there are correlations between the in and the out states. The out and in regions are separated by the horizons. The degrees of freedom on or close to the horizon evolve, entering with the ingoing particles and information leaves with the outgoing particles. Assuming a minimal locality in the laws of nature the degrees of freedom which are evolving should be localized near the horizon. Hence, taking into account this preference, we rewrite Eq.~(\ref{newgenSfactor}) in the following manner,
\begin{align}
S(g_1+g_2) = S_{\text{out}}S_{\text{horizon}}S_{\text{in}}, \nonumber \\
S_{\text{out}} = e^{i\phi(g_1,g_2)}S_2(g_2)e^{-i\phi(g_1,g_2)};~~S_{\text{horizon}}  = e^{i\phi(g_1,g_2)};~~S_{\text{in}} = S_1(g_1).
\label{thooftfactor}
\end{align}
The work of 't Hooft also gives us confidence in this identification. There the in and out particles certainly interact by gravitational forces and 't Hooft \cite{thooft1,stephens,thooft1.5,thooft2,thooft2.5}, taking into account the effect of frame dragging finds a S matrix of a similar form as above which is governed by local commutation rules. The applications of our approach in section~(\ref{Applications}) also provide conclusions which support this identification. We should point out that this is not the only possibility. In fact, one example we will address in section~(\ref{FKconnection}) involves finding the asymptotic states in the theory of infrared divergences in gauge and gravitational theories with massless particles at the perturbative level. Here the infrared degrees of freedom do live on surfaces at $t =\pm \infty$. The corresponding expression for $\phi$ is calculable and is highly nonlocal. This again is not unexpected since in this situation we are only interested in the infrared divergent long distance effects.

At this point we would like to carefully provide the definitions of the various factors involved in Eq.~(\ref{thooftfactor}) for the case of the quantum black hole. These definitions can be best understood by imagining a sphere of radius say, $r=3M$ around the black hole ($M$ is the mass of the black hole) and then describing the particles flowing in and out of this sphere. $S_{in}$ is the ``in" S matrix associated with the the objects flowing in asymptotically outside the sphere.
$e^{i\phi(g_1,g_2)}$ is the horizon S matrix, $S_{\text{hor}}$ which describes how particles from the sphere scatter against the black hole and then reemerge out of the sphere. As the dependence of $\phi$ on both $g_1$ and $g_2$ indicates, it describes how the inward moving particles close to the horizon interact with the outward moving ones. As we will see, this factor is the most important one of the total S matrix. What happens from there on till the particles reach future infinity is described by the ``out" S matrix $S_{\text{out}}$. In the above derivation of Eq.~(\ref{newgenSfactor}) and earlier of Eq.~(\ref{newSfactor}) we have assumed that the three S Matrices are by themselves unitary. In addition to what we have said about this earlier, we note that while this assumption is quite reasonable for the primary contributions involving the flow of light particles in and out, the question of the exact unitarity of the separate factors by themselves would require more knowledge of the full quantum theory of gravity. If this assumption is not supported by such a theory then the factorization formula may need some modification. We will return to additional explanations of this point in section (\ref{causalityeft}). However, we should emphasize that the evolution equation we derive below only uses Eq.~(\ref{newgendynamics2}).

An important point we must address now is the conditions under which we may have causality violation at any stage of the evolution of the S matrix. As we will discuss in the next subsection, in order to discuss the evolution in terms of a time $\tau$, we must change change our infinitesmal description from $\delta g(x)$ to $\delta\tau$. The prescription for this is $\delta g(x) = f^{\prime}(\tau-x^0)\delta \tau$, where $f^{\prime}(\tau-x^0)$ is essentially a $\delta$ function \cite{Bogoliubov} (Please see the discussion after Eq.~(\ref{Schrodinger})). Thus in general if $\phi$ is time independent or if its time dependence may be ignored, then from Eq.~(\ref{newgendynamics2}) there will not be any causality violation. This is the case for the Hawking calculation, where we will explicitly determine the time independent $\phi$ (see Eq.~(\ref{bogoliubov-tr})) expressed in terms of the creation and annihilation operators. It is also true for the eternally accelerating observer of the Unruh effect. Moreover, the causality violation is restricted to the region where the horizon S matrix is defined. It is just that between the in and out regions where the inward moving particles close to the horizon interact with the outward moving ones. In other words, it is non-zero only at the horizon when the horizon hamiltonian defined in Eq.~(\ref{horizon-H}) is time dependent.

Let us briefly consider the nature of the transformation, Eq.~(\ref{genginvariance}). We have seen that such a transformation does not change the unitarity condition on the S matrix. However, it does change the causality condition and through it the nature of the Hilbert space of the theory. In section~(\ref{complementarity}) we will consider an interpretation of these transformations and find that under these the S matrix transforms covariantly. In this sense they do not leave the entire theory invariant since the transformed theory has a different Hilbert space structure than the original. They can be thought of as symmetries only in an ``enlarged'' Hilbert space. By the ``enlarged'' Hilbert space we mean appending that obtained by applying the operation Eq.~(\ref{genginvariance}). We will see later that the ``symmetry" Eq.~(\ref{genginvariance}) is spontaneously broken. It is precisely because of the above remarks that no Goldstone bosons will be associated with this breaking. 

Incidentally, it is worth noting that a proposal for resolving the unitarity problem of black hole evaporation  was made in \cite{maldacena} where a final state boundary condition was imposed at the singularity. This proposal requires a specific final state which is perfectly entangled between the infalling matter and the incoming Hawking radiation. Our formulation is general enough to include this situation as well. In fact, as we argue in section~(\ref{complementarity}), instead of thinking of $e^{i\phi}$ as the horizon S matrix we can use it to implement the correlations between the infalling matter and Hawking radiation needed for the realization of the proposal of \cite{maldacena}.
%%%%%%%%%%%%%%%%%%%%%%%%%%%%%%%%%%%%%%%%%%%%%%%%%%%%%%%%%%%%%%%%%%%%%%%%%%%%%%%%%%%
\subsection{An Evolution Equation for the S Matrix}
\label{evolution}
Next, let us write down a solution of Eq.~(\ref{newgendynamics2}), i.e., let us now consider how the new dynamics, with $\phi$ taken into account, changes the solution for the new S-matrix. Formally this solution can be written in the future region (out states) as,
\begin{equation}
i{\delta S'(g) \over \delta g(x)}S'^{\dagger} =  i{\delta e^{i\phi}  \over \delta g(x)}e^{-i\phi} +  e^{i\phi} H(x,g)e^{-i\phi},
\label{futuredynamics}
\end{equation}
where we have appropriately identified  $H(x,g)$ with $i{\delta S(g) \over \delta g(x)}S^{\dagger}$. It is the generalized total Hamiltonian for the case when $\phi = 0$ since $H(x,g)$ satisfies the usual causality condition (see Eq.~(\ref{Causality1}) and Eq.(\ref{hcausality}) below). We note that we are using the boundary conditions Eq.~(\ref{bc}). The solutions that we are interested in are only those where the $\phi$ dependance of the left hand side Eq.~(\ref{newgendynamics2}) is cancelled amongst the two terms there and trivial solutions with $g_2=0$ are ignored. There is an exactly symmetrical equation for the past region (incoming states). These equations again reflect the same situation as Eq.~(\ref{newgenSfactor}), i.e., the in and out states are not independent of each other. 

Before we write down a formal solution, we will take a short digression to see how the notion of time and a Schrodinger  equation appear even for the standard situation in flat space without introducing $\phi$.
The equation for the S matrix in this case may be written as 
\begin{equation}
i\frac{\delta S(g)}{\delta g(x)} = H(x,g)S(g)
\label{Schrodinger}
\end{equation}
 or in the integral form as,
 \begin{equation}
i\delta S(g) = \int H(x,g)S(g)\delta g(x) d^4x.
\label{schrodinger}
\end{equation}
We recall here that the generalized Hamiltonian $H(x,g)$ can be written as an expansion,
\begin{equation}
H(x,g) = H(x) + \sum_{n \geq 1}\frac{1}{n!}\int H_n(x,x_1,...,x_n)g(x_1)...g(x_n)d^4x_1...d^4x_n,
\label{hexpansion}
\end{equation}
and from Eq.~(\ref{constraint1}), the causality requirement on the S matrix implies,
\begin{eqnarray}
H_n(x,x_1,...,x_n) = 0, ~~n=1,.... \nonumber \\
\text{if for at least one}~ x_j (j=1,...), x^0 > x_j^0.
\label{hcausality}
\end{eqnarray}
Thus, the integration over the $x_i$ in Eq.~(\ref{hexpansion}) takes place over the future directed light cone at $x$. Because of this, for $g(x)$ one may choose a smooth function that falls off rapidly at large $x_j^0$. A suitable choice (this discussion is from \cite{Bogoliubov}) is given by,
\begin{eqnarray}
g(x) = f(\tau-x^0) \nonumber \\
f(\tau-x^0) = \theta (\tau-x^0), \text{for}~|\tau-x^0| \geq \Delta t,
\label{gee}
\end{eqnarray}
which corresponds to the situation when the interaction is switched on for times $-\infty$ to $\tau - \Delta t$ and slowly switched off from $\tau - \Delta t$ to $\tau + \Delta t$. It is in this way that the usual causal ordering is introduced. The operator $S$ is now a function of $\tau$ and using the fact that $\delta g(x) = f^{\prime}(\tau-x^0)\delta \tau$, we get,
\begin{equation}
i\frac{\delta S(\tau)}{\partial \tau} = \int H(x,f)f^{\prime}(\tau-x^0)d^4xS(\tau)
\label{semiclassical}
\end{equation}
The integration in Eq.~(\ref{hexpansion}) is over the light cone of height $\tau + \Delta t - x^0$ and since $x^0 > \tau - \Delta t$, we see that the integration in Eq.~(\ref{hexpansion}) is along the region:
\begin{equation}
|x_j^0 - x^0| \leq 2\Delta t~~;~~ |{\mathbf{x_j}} - {\mathbf{x}}| \leq 2\Delta t.
\end{equation}
Thus, the generalized Hamiltonian only depends on the dynamical variables in the neighborhood $2\Delta t$ of $x$. If the limit $\Delta t \rightarrow 0$ can be taken then we have the Schrodinger equation in the interaction representation,
\begin{equation}
i\frac{\delta S(\tau)}{\partial \tau} = \int H(x)d^3xS(\tau)
\end{equation}
This discussion can be generalized to the case when the switching off of the interaction is not along the plane $x^0=\tau$ but along a space-like surface $\sigma$. Here we make the replacement, as $\Delta t \rightarrow 0$,
$g(x) \rightarrow \theta_{\sigma}(-x) = \sigma(x)$ to obtain,
\begin{equation}
i\frac{\delta S(\sigma)}{\delta \sigma} = H(x,\sigma)S(\sigma).
\label{Tomonaga}
\end{equation}
The argument above for obtaining the Schrodinger equation clearly depends on the limit $\Delta t \rightarrow 0$ existing. However, there are many instances where this is not possible. The most obvious situation is when there are infrared divergences in the theory. This problem, however, can be resolved using the Faddeev-Kulish \cite{FK} method. This resolution is a special case application of the methods discussed in this section and is taken up in some detail in the next section. Another instance where the limit cannot be taken is for the case of vacuum diagrams in higher order of perturbation theory where one has counter-terms proportional to
$c ~\Box \delta^4(x-x^{\prime})$. In this case the generalized Hamiltonian has the contribution:
\begin{equation}
c{\partial^2 \over \partial x^2} \int \delta^4(x-x^{\prime}) g(x^{\prime}) d^4x^{\prime}= c {\partial^2 g(x) \over \partial x^2}.
\end{equation}
If we use Eq.~(\ref{gee}) then we have a contribution to the generalized Hamiltonian which is not integrable. Clearly this means that the theory does not like to be local. The process of the switching off the interaction has a very dramatic effect on the system such that all previous memory is erased. Note that this problem exists regardless of whether the constant $c$ above is divergent or finite. These problems are of great consequence for the cosmological constant problem. It appears that the S matrix approach presented here offers at least some relief by working with smooth functions $g(x)$. The flip side is that the concept of time is not transparent though we can consider $g(x)$ to be clock functions. Perhaps this is the price to pay for the resolution of the cosmological constant problem. The erasure of previous memory if we insist on the introduction of time through the switching off procedure is suggestive that an uncertainty principle is at work here. This ends our digression.

When we apply the above discussion to the case being considered in this section, i.e., the description of systems which require that $\phi \neq 0$ then the above problems become more immediate. Indeed, here the generalization of 
Eq.~(\ref{schrodinger}) from Eq.~(\ref{futuredynamics}) is,
\begin{equation}
i\delta S(g) = \int i\frac{\delta e^{i\phi}}{\delta g(x)}e^{-i\phi} S(g) \delta g(x) [d^4x] + \int e^{i\phi}H(x,g)e^{-i\phi}S(g)\delta g(x) [d^4x].
\label{integratedfuturedynamics}
\end{equation}
Note that $H(x,g)$ still satisfies the condition, Eq.~(\ref{hcausality}) on account of the causality condition  
Eq.~(\ref{causality-prime}). {\emph{Hence the integration in the second integral is again restricted to the future directed light cone at $x$}}. Thus, it is important to realize that this second term is amenable to semiclassical analysis even though it contains information about the horizon degrees of freedom through the factors of $e^{i\phi}$. The first integral involves the horizon degrees of freedom and the integration here, however, is unrestricted from causality considerations. The quantity $i\frac{\delta e^{i\phi}}{\delta g(x)}e^{-i\phi}$ is just the generalized horizon Hamiltonian $H_h(x,g)$. The boundary S matrix then is seen to dress or clothe the hamiltonian $H(x,g)$. Recall that this is just the hamiltonian in the absence of the entangling unitary 
$e^{i\phi}$. 

The generalized horizon Hamiltonian, $H_h(x,g)$, may be expanded as: 
\begin{equation}
H_h(x,g) = \int [d^4x_1]H^{(1)}_h(x,x_1)g(x_1) + \int [d^4x_1] \int [d^4x_2] H^{(2)}_h(x,x_1,x_2)g(x_1)g(x_2) + ...
\label{phiexpansion}
\end{equation}
It should be noticed that it contains knowledge of the entire history which is relevant  to the transfer of the information from the in to the out states. This aspect is very similar to the abelian example discussed in the appendix. It is important to note that because of the entangling boundary conditions there is no term in the expansion in Eq.~(\ref{phiexpansion}) at zeroth order in $g(x)$. Thus there is no analog of $H(x)$ in the expansion, Eq.~(\ref{hexpansion}). In addition $H_h(x,g)$ does not satisfy any causality condition and so no condition like Eq.~(\ref{hcausality}) can be used to essentially reduce it to a  local term. This is an entirely new type of nonlocal object that enters into the analysis and does not contribute in the semiclassical limit (see the discussion below and the next section).
The first term on the RHS of Eq.~(\ref{integratedfuturedynamics}) is responsible for the dynamics of the horizon S matrix. It can be ignored when considering equilibrium situations like the eternal black hole or a black hole in equilibrium with the radiation. In this way the Hawking result is protected from any controversy. However, any possible evidence for firewalls \cite{Firewalls,Firewalls1} would appear in an analysis of this first term. In addition, it is exactly this term in the evolution of the S matrix which is responsible for the consequences of causality violation. These consequences of the violation of Bogoliubov causality will be discussed further in section \ref{causalityeft}.

Summarizing, the infinitesimal evolution of the S matrix can be written as,
\begin{equation}
i\delta S(g) = \int H_h(x,g)S(g) \delta g(x) [d^4x] + \int e^{i\phi}H(x,g)e^{-i\phi}S(g)\delta g(x) [d^4x].
\label{theequation}
\end{equation}
This equation expresses the evolution of total S matrix of Eq.~(\ref{newgenSfactor}) or Eq.~(\ref{thooftfactor}). The concept of a universal time may be introduced (with all its associated problems) for the integral in the second term. In the first term the integration is not restricted to the forward light cone and is therefore the perpetrator of causality violation.
%%%%%%%%%%%%%%%%%%%%%%%%%%%%%%%%%%%%%%%%%%%%%%%%%%%%%%%%%%%%%%%%%%%%%%%%%%%%%%%%%%%
\section{Some Dynamical Applications of the New S matrix Approach}
\label{Applications}
In the previous section we have introduced a new S matrix approach to study the dynamics of systems which exhibit correlations between the in and out states in the presence of the boundary S matrix. Our discussion so far has been formal and it would be good to see some of the consequences in explicit examples. However, a completely satisfactory analysis with a well defined approximation scheme is still lacking beyond those which will be discussed below. However, we would like to argue with some examples that much information about the quantum black hole may be obtained from general considerations. Our construction of the evolution equation is such that the S matrix is manifestly unitary, however, the consequences of the violation of Bogoliubov causality must be properly analyzed to determine if there is any information loss due to it. For now we will assume that in spite of the causality violation a basis may be found where a consistent causal ordering results.
Both terms in Eq.~(\ref{integratedfuturedynamics}) are crucial in implementing these aspects of unitarity. As discussed above $H(x,g)$ denotes the generalized Hamiltonian in the absence of a horizon. Thus $e^{i\phi}H(x,g)e^{-i\phi}$ expresses the role of the horizon S matrix as a ``clothing" or ``dressing" operator as far as the outgoing states are concerned. In this way we see that the quantum state of the outgoing particles is determined by the horizon S matrix which involves all the interactions, including gravitational and matter interactions. We will first outline how the standard Hawking result may be reproduced by ignoring the effect of the quantum gravitational interactions at the horizon and keeping only the second term on the RHS of Eq.~(\ref{theequation}). Then we include the gravitational effects in the eikonal approximation following the work of 't Hooft. We briefly review how this introduces the interactions between the in and the out states and in this way discuss how the true vacuum with correlations between the quanta in the regions \Rmnum{1} (and \Rmnum{3}) and \Rmnum{2} (and \Rmnum{4}) of the Kruskal manifold may be obtained from the Hawking vacuum. After that we will discuss the implications for the ground state of quantum gravity. We argue that the choice of entangling boundary conditions Eqs.~(\ref{bc}) allow us to write down the form of the ground state, at the horizon, in the full quantum gravity theory.

In section~(\ref{complementarity}) we will provide an interpretation of the transformation Eq.~(\ref{genginvariance}) and also explain how the final state projection models of \cite{maldacena} are easily accommodated into our formulation. In the following section~\ref{causalityeft} we will further explore the consequences of Eq.~(\ref{integratedfuturedynamics}) with regard to causality and the effectiveness of field theoretical descriptions. 

We have seen that one consequence of the generalized causality is that the in and out states depend on the generalized coupling parameter $g(x)$. An example of where the coupling cannot be switched off is the case of long range interactions which give rise to infrared divergences. This is a case that has been studied previously, \cite{FK} and the change of the Hilbert space structure discussed in the previous section is explicitly realized in the asymptotic state dynamics of Faddeev and Kulish. Since the black hole case discussed has many novel features it is worth seeing how the formalism works out for the case of infrared divergences in QED and in perturbative quantum gravity where the evolution equation, Eq.~(\ref{integratedfuturedynamics}) can solved in the soft approximation. This is done in sub-section \ref{FKconnection}.  The infrared sensitive part of the S matrix can be completely evaluated in this perturbative approach and the Hilbert space structure can be explicitly analysed. Our conclusions are in agreement with earlier work \cite{FK} and provide confidence in the validity of Eq.~(\ref{integratedfuturedynamics}).
%%%%%%%%%%%%%%%%%%%%%%%%%%%%%%%%%%%%%%%%%%%%%%%%%%%%%%%%%%%%%%%%%%%%%%%%%%%%%%%%%%%
\subsection{Hawking Radiation, Including Gravitational Interactions and the Ground State of Quantum Gravity}
\label{Hawking}
As mentioned at the end of the previous section, it is the second term on the RHS of Eq.~(\ref{integratedfuturedynamics}) that is amenable to semiclassical analysis and from this the standard results for Hawking radiation are reproduced. A justification for the neglect of the first term can be provided by limiting ourselves to incoming modes near the horizon which are of low energies. In the first term we can then make the changes discussed between Eq.~(\ref{hcausality}) and Eq.~(\ref{Tomonaga}) of the previous section and then it is seen that time derivatives are small. With the same changes implemented into the evolution equation, it may be written as:
\begin{align}
i{\delta S(\sigma) \over \delta \sigma} = \widetilde{H(x,\sigma)} S(\sigma) \nonumber \\
 \widetilde{H(x,\sigma)} = e^{i\phi} H(x,\sigma) e^{-i\phi}
 \label{semiclassical}
\end{align}
Note that $H(x,g)$ is the full generalized Hamiltonian when there is no horizon present. The standard Hawking result can be obtained by only keeping the quasi-free Hamiltonian $H_0$ on the RHS of Eq.~(\ref{semiclassical}). However from the standpoint of this paper it is clear that interactions are easily incorporated. Correlations are later introduced for example by including the quantum gravitational contributions to the horizon S matrix $e^{i\phi}$ but in the leading semiclassical approximation, $\phi$ just plays the role of the generator of the Bogoliubov transformation. Let us discuss this point in some more detail.

Consider a Kruskal manifold, though in the near horizon limit, the Rindler coordinates would also be appropriate. Then, as discussed earlier, we have, $\phi = \phi(g_{\Rmnum{1}}(x), g_{\Rmnum{2}}(x))$, where $g_{\Rmnum{1}}(x)=1$, if $x$ is in region \Rmnum{1} and is zero otherwise and similarly for $g_{\Rmnum{2}}(x)$ in region \Rmnum{2}. It is important to remember for the following discussion that $\phi$ satisfies the boundary conditions, Eq.~(\ref{bc}), i.e., 
\begin{equation}
\phi(g_{\Rmnum{1}},0) = \phi(0,g_{\Rmnum{2}}) = 0~;~ \phi(0,0) = 0.
\label{bcnew}
\end{equation}
The Hamiltonian $H(x,g)$ in the second term of Eq.~(\ref{integratedfuturedynamics}) is the one without any reference to the horizon and it is composed of field operators $\Phi(x)$
for a scalar field on the Kruskal manifold which can be expanded in terms of the normal mode solutions of the Klein-Gordon equation (we follow the notations of \cite{Israel} here, see also \cite{dewit}),
\begin{equation}
[F_{\omega j}(x)]_{\sigma} = \frac{1}{(2\pi)^{1/2}(2\omega)^{1/2}r}e^{-i\omega t}R_{\omega l\sigma}(r)Y_{lm}(\theta,\varphi),
\end{equation}
where, the energy, $\omega \geq 0$ and $j$ denotes $(l.m)$. In the above, $\sigma$ denotes the two regions \Rmnum{1}, and \Rmnum{2}. The expansion for the scalar field is,
\begin{align}
\Phi(x) = \sum_{\omega j}\left[a_{\omega,j}^{\Rmnum{1}}[F_{\omega j}(x)]_{\Rmnum{1}} + {a^{\dagger}}_{\omega j}^{\Rmnum{1}}[F^{*}_{\omega j}(x)]_{\Rmnum{1}}\right] \nonumber \\
+\sum_{\omega j}\left[{a^{\dagger}}_{\omega j}^{\Rmnum{2}}[F_{\omega j}(x)]_{\Rmnum{2}} + a_{\omega j}^{\Rmnum{2}}[F^{*}_{\omega j}(x)]_{\Rmnum{2}}\right]
\end{align}
The summation includes an integration over the continuous variables.
The sets of creation and annihilation operators $\{a_{\omega j}^{\Rmnum{1}}, {a^{\dagger}}_{\omega j}^{\Rmnum{1}}\}$ and  
$\{a_{\omega j}^{\Rmnum{2}}, {a^{\dagger}}_{\omega j}^{\Rmnum{2}}\}$ are those for the parochial observers in the two regions respectively and the non-vanishing commutation relations are,
\begin{equation}
[a_{\omega j}^{\sigma}, ~ {a^{\dagger}}_{\omega' j'}^{\sigma'}] = \delta_{\sigma \sigma'}\delta_{j j'}\delta(\omega-\omega^{\prime}).
\end{equation}
The scalar field may also be expanded in terms of the modes $H_{\omega j}$ and ${\tilde{H}}_{\omega j}$ which are analytic in the entire Kruskal manifold. From analyticity considerations \cite{Israel,unruh}, these modes are related to the earlier ones by,
\begin{equation}
H_{\omega j} = [F_{\omega j}]_{\Rmnum{1}} \cosh \lambda_{\omega} + [F_{\omega j}]_{\Rmnum{2}} \sinh \lambda_{\omega},
\end{equation}
and a similar relation with \Rmnum{1} replaced by \Rmnum{2} for ${\tilde{H}}_{\omega j}$. The analyticity requirement is that $\tanh \lambda_{\omega} = e^{{-\pi \omega \over \kappa}}$, where $\kappa$ is the surface gravity.
In terms of these modes the scalar field has the expansion, 
\begin{align}
\Phi(x) = \sum_{\omega,j}\left[a_{\omega j}(\kappa)H_{\omega j}(x) + {a^{\dagger}}_{\omega j}(\kappa)H^{*}_{\omega j}(x)\right] \nonumber \\
+\sum_{\omega j}\left[{{\tilde{a}}^{\dagger}}_{\omega j}(\kappa){\tilde{H}}_{\omega j}(x) + {\tilde{a}}_{\omega j}(\kappa){\tilde{H}}^{*}_{\omega j}(x)] \right].
\end{align}
The sets of creation and annihilation operators $\{a_{\omega j}(\kappa), {a^{\dagger}}_{\omega j}(\kappa)\}$ and  
$\{{\tilde{a}}_{\omega j}(\kappa), {{\tilde{a}}^{\dagger}}_{\omega j}(\kappa)\}$ are those relevant for the Kruskal vacuum $\left|0(\kappa)\right>$, which is the physical vacuum associated with the space-like hypersurface $t =$ constant. The non-vanishing commutation relations are,
\begin{equation}
[a_{\omega j}(\kappa), ~ {a^{\dagger}}_{\omega j'}(\kappa)] = [{\tilde{a}}_{\omega j}(\kappa), ~ {{\tilde{a}}^{\dagger}}_{\omega j'}(\kappa)] = \delta_{j j'}\delta(\omega-\omega').
\end{equation}
From the second term of Eq.~(\ref{integratedfuturedynamics}), it is easily seen that the horizon S matrix which implements the Bogoliubov transformation i.e.,
\begin{align}
a_{\omega j}(\kappa) = e^{i\phi} a_{\omega j} e^{-i\phi},
\end{align}
etc, in leading semiclassical approximation is given by:
\begin{equation}
\phi = - \sum_{\omega,j} i\lambda_{\omega}({a^{\dagger}}_{\omega j}^{\Rmnum{1}}{a^{\dagger}}_{\omega j}^{\Rmnum{2}} - a_{\omega j}^{\Rmnum{1}}a_{\omega j}^{\Rmnum{2}}).  
%= - \sum_{\omega,j} i\lambda_{\omega}({\tilde{a}}^{\dagger}_{\omega j}(\kappa){{\tilde{a}}^{\dagger}}_{\omega j}(\kappa) - {\tilde{a}}_{\omega,j}(\kappa){\tilde{a}}_{\omega j}(\kappa)
\label{bogoliubov-tr}
\end{equation}
An obvious feature of this equation which is worth emphasizing is the symmetric manner in which the two regions $\Rmnum{1}$ and $\Rmnum{2}$ appear. Indeed, if we write $\phi(\alpha^{\Rmnum{1}},\alpha^{\Rmnum{2}})$, where $\alpha = a_{\omega j}$ or $a_{\omega j}^{\dagger}$, then Eq.~(\ref{bogoliubov-tr}) satisfies $\phi(\alpha^{\Rmnum{1}},0) = \phi(0,\alpha^{\Rmnum{2}}) = 0$. This again is a reflection, even at this level, of the imposition of the entangling boundary conditions, Eq.~({\ref{bc}}).
If, $\left|0^{\Rmnum{1}} 0^{\Rmnum{2}}\right>$  denotes the vacuum of the parochial observers in regions \Rmnum{1} and 
\Rmnum{2}, then the physical vacuum experienced by the freely falling observer is, using Eq.~(\ref{bogoliubov-tr}),
\begin{align}
\left|0(\kappa)\right> = e^{i\phi}\left|0^{\Rmnum{1}} 0^{\Rmnum{2}}\right> \nonumber \\
= \prod_{\omega}(\cosh \lambda_{\omega})^{-1}e^{(e^{-\pi\omega/\kappa}\sum_j{a^{\dagger}}_{\omega j}^{\Rmnum{1}}{a^{\dagger}}_{\omega j}^{\Rmnum{2}})}\left|0^{\Rmnum{1}} 0^{\Rmnum{2}}\right>
\label{Kvacuum1}
\end{align}
Straightforward algebra then gives,
\begin{equation}
\left|0(\kappa)\right> = \prod_{\omega,j}(\cosh \lambda_{\omega})^{-1}\sum_{n_{\omega j}=0}^{\infty} e^{-n_{\omega j}\pi\omega/\kappa}\left|n_{\omega j}^{\Rmnum{1}} n_{\omega j}^{\Rmnum{2}}\right>.
\label{Kvacuum2}
\end{equation}
From this we see that observers outside the horizon perceive a thermal spectrum and the usual problems with unitarity arise. The Kruskal vacuum is often called the Hawking vacuum.

At the semi-classical level it is now seen that the symmetry of equation Eq.~(\ref{genginvariance}) is spontaneously broken. 
%First, we note that  the free Hamiltonian is given by \cite{Israel, thooft-rev},
%\begin{align}
%H_0 = [H_0]_{\Rmnum{1}} - [H_0]_{\Rmnum{2}} \nonumber \\
%= \sum_{\omega j}\omega {a^{\dagger}}_{\omega j}^{\Rmnum{1}}a_{\omega j}^{\Rmnum{1}} - 
%\sum_{\omega j}\omega{a^{\dagger}}_{\omega j}^{\Rmnum{2}}a_{\omega j}^{\Rmnum{2}}.
%\end{align}
%The expression for $\phi$ given in Eq.~(\ref{bogoliubov-tr}) commutes with the $H_0$, i.e., $[H_0 , \phi] = 0$. 
Indeed, it follows from the above discussion that $\phi\left|0(\kappa)\right> \neq 0$, thus the physical vacuum is not invariant under the symmetry. There are no goldstone bosons associated with this because as mentioned earlier the symmetry changes the Hilbert space structure and does not leave the whole theory invariant. 

An obvious feature of Eq.~(\ref{Kvacuum2}) is that the quanta in the two regions are uncorrelated, which of course is ultimately responsible for the thermal nature of the Hawking radiation which in turn drives the Hawking information loss argument. How does the S matrix unitarize itself? In the above analysis we have ignored gravitational interactions, in particular we have ignored any quantum gravitational backreaction on the final state of the outgoing radiation. This is consistent with Hawking's original calculation and his point of view in \cite{Hawking} that strong gravitational effects are important at later times and far inside the horizon as the singularity is approached. A diametric view has been advocated by 't Hooft who has argued precisely for strong gravitational effects near the horizon (from the point of view of an outside observer) between the incoming (regions \Rmnum{2} in the Kruskal manifold)and the outgoing states (regions \Rmnum{1} of the Kruskal manifold) . In particular particles coming in very early and those leaving the black hole very late have an exponentially enhanced center of mass energy which implies an out of control gravitational interaction. Thus the horizon is a gravitationally strongly interacting region for an outside observer. In a series of papers, 't Hooft has proposed an S matrix ansatz to study precisely whether these strong interaction effects can resolve the unitarity problem. At first only purely gravitational interactions are considered in the context of the eikonal approximation and the unitary horizon S matrix in the splitting Eq.~(\ref{thooftfactor}) is calculated. The justification for this approximation is the following. We are considering the interaction of the infalling and outgoing particles near the horizon with huge center of mass energies. However, the scattering takes place near the horizon at different directions and hence at small angles. Thus, we are in the kinematic regime which is well approximated by large center of mass energy ($\sqrt{s}$) and small momentum transfer ($t$) in which graviton exchanges are dominant over any other string effects. We will discuss the unitarization mechanism in this approximation since the corresponding horizon S matrix for pure gravity has been explicitly calculated by 't Hooft \cite{thooft1,stephens,thooft1.5}. First we note that in this approximation the  graviton exchanges carry low momentum compared to the matter projectiles. Even if the paths of the particles are not straight, the graviton effects will factorize from the matter and we can write the horizon S matrix $e^{i\phi}$ in the second term on the RHS of Eq.~(\ref{integratedfuturedynamics}) in the following form:
\begin{equation}
e^{i\phi} \approx (e^{i\phi})_{grav}(e^{i\phi})_{matter}.
\end{equation}
Thus in the second term in the RHS of Eq.~(\ref{integratedfuturedynamics}), we now have,
\begin{equation}
e^{i\phi} H(x,g) e^{-i\phi} \rightarrow (e^{i\phi})_{grav}[(e^{i\phi})_{matter} H(x,g) (e^{-i\phi})_{matter}](e^{-i\phi})_{grav}.
\end{equation}
It is only the term inside the square brackets which is calculated above in the Hawking approximation. As we will argue below, the gravitational contribution to the horizon S matrix will change the nature of the Bogoliubov transformation and introduce correlations between the modes in regions \Rmnum{1} and \Rmnum{2}. These correlations make the process unitary. We would like to note that the true vacuum in these circumstances is no longer the Hawking vacuum, $\left|0(\kappa)\right>$, but,
\begin{equation}
\left|0\right> = (e^{i\phi})_{grav}\left|0(\kappa)\right>.
\end{equation}
This vacuum embodies the correlations between the in and the out states. In an attempt to explain how these correlations arise, we will next briefly review the 't Hooft construction\cite{thooft1,stephens,thooft1.5} for $(e^{i\phi})_{grav}$.  In principle this can also be obtained from Eq.~(\ref{integratedfuturedynamics}) by using the Einstein action and working in the eikonal approximation. Before we describe the 't Hooft construction we briefly remark that in the near horizon limit for massive black holes the Kruskal coordinates may be approximated by the Rindler coordinates with the replacement of $(\theta,\varphi) \rightarrow {\tilde{x}} = (x,y)$. Following 't Hooft we will use these coordinates in the following.

In the eikonal approximation the gravitational interactions are described by shockwaves \cite{thooft-dray}. As they approach the horizon, the incoming particles experience a large Lorentz boost causing a frame dragging effect with a delta function support there. The net effect is a shift in the position (more accurately the shape) of the horizon. Since transverse momenta are small, it is assumed that states may be specified by the longitudinal momenta as a function of the transverse position. If the incoming momentum are in the $-$ direction and the outgoing in the $+$, 't Hooft finds that the S matrix which is time reversal invariant may be written as,
\begin{align}
\langle \{P_{out}^-({\tilde{x}})\}\left|\{P_{in}^+({\tilde{x}}')\}\right> =  e^{-i\int d^2{\tilde{x}}d^2{\tilde{x}}'P_{in}^-({\tilde{x}}')f({\tilde{x}} - {\tilde{x}}')P_{out}^+({\tilde{x}})}. \nonumber \\
{\tilde{\partial^2}}f = - \delta^2({\tilde{x}} - {\tilde{x}}'). 
\label{thooft-Smatrix}
\end{align}
Here, $P_{out}^+({\tilde{x}})$ is the quantum generator of the shift at ${\tilde{x}}$ and it also measures the {\emph{total}} outgoing momentum there, and $P_{in}^-({\tilde{x}})$ is the same for the incoming states. In order to see that this implies highly non-trivial correlations between the ingoing and outgoing particles let us defined the canonically conjugate operators $X_{in}^+({\tilde{x}})$ and $X_{out}^-({\tilde{x}})$. Then it can be shown that 
\begin{equation}
[X_{out}^-({\tilde{x}}) , X_{in}^+({\tilde{x}}')] = if({\tilde{x}} - {\tilde{x}}').
\end{equation}
Physically, $X_{out}^-({\tilde{x}})$ may be interpreted as indicating the position of the horizon with respect to the outgoing particles and $X_{in}^+({\tilde{x}})$  is associated with the position of the past horizon with respect to the incoming particles. It should be emphasized here that the gravitational shift interaction considered by 't Hooft shows how transfer of information about one particular quantum number, namely the longitudinal momentum can take place. In the eikonal approximation used in the calculation, the longitudinal modes are strongly coupled but the transverse ones are treated semiclassically. Thus the problems with causality mentioned earlier should not be an issue since only radially incoming and outgoing directions are involved and no closed loops of information flow can form. We will return to a general discussion of this point in section \ref{causalityeft}. Finally it is worth pointing out that the 't Hooft construction is valid only when the transverse separation between the particles is much bigger than the planck length. To do better in his approach one will have to make additional assumptions about the interactions at the planck scale. However as we discuss below, the symmetry of the horizon S matrix between the observables in the incoming and outgoing regions will remain in the exact theory.

In concluding this discussion, we emphasize an important point that the horizon S matrix is off-diagonal in the quantum numbers of the in and out states. This is explicit in the 't Hooft form, Eq.~(\ref{thooft-Smatrix}). More generally without any approximations, we had argued earlier that  our $\phi(g_1,g_2)$ is symmetric under the interchange $g_1\leftrightarrow g_2$ and satisfies the boundary conditions Eq.~(\ref{bc}). The 't Hooft result, Eq.~(\ref{thooft-Smatrix}), is a thus the only possible outcome for the horizon S matrix that can correlate the incoming and outgoing longitudinal momenta and, again, automatically reflects the choice of entangling boundary conditions, Eq.~(\ref{bc}), that we have been advocating. Including all the interactions in the horizon S matrix would in principle ensure entanglement between the ingoing and the outgoing bits, and our choice of entangling boundary conditions will ensure the symmetry between them in the corresponding expression for $\phi(g_1,g_2)$, however complicated it may be. That this mixing, however small, can lead to unitarization is pointed out in \cite{page,RajuPapa}. In summary, in the absence of a complete calculation, let us organize the various steps as seen from the point of view of the RHS of Eq.~(\ref{integratedfuturedynamics}). First we start with the free generalized Hamiltonian included in $H(x,g)$, then we use the semiclassical approximation to dress it with the horizon S matrix which leads to the purely thermal Hawking vacuum. Next we take the first term into account and include all the interactions in the {\emph{ off diagonal horizon S matrix}} including gravitational, which produces correlations between the incoming and outgoing radiation that eventually unitarizes the whole process. 
We are of course assuming here that the causality violation is temporary in the sense that outside the horizon, a standard causal ordering may be realized. We will return to a discussion of this point in section (\ref{causalityeft}).

We will now argue that the structure of the vacuum of the full theory not just in the semiclassical approximation, has a form similar to Eq.~(\ref{Kvacuum2}). We have seen that it is the horizon S matrix which completely determines the quantum state of the outgoing matter. In particular the physical out vacuum, including all possible interactions  in $\phi$ is given by:
\begin{equation}
\left|0\right> = e^{i\phi}\left|0_L 0_R\right>,
\end{equation}
where we are using the Rindler coordinates in the near horizon limit and $\left|0_L 0_R\right>$ is the uncorrelated vacuum of the parochial observers in the two wedges. The off diagonal horizon S matrix introduces strong correlations between the two wedges which perhaps even replaces the horizon and perhaps even a geometrical description of spacetime becomes inappropriate. However, because of the structure of the horizon S matrix which is symmetric in the disturbances in the two wedges due to the choice of the entangling boundary conditions, Eqs.~(\ref{bc}, \ref{bcnew}), the physical out vacuum must still be expressible in the form,
\begin{equation}
\left|0\right> = \sum_i C_i(\omega_i, \gamma)\left|\omega^L_i\right> \otimes \left|\omega^R_i\right>.
\end{equation}
Here $\left|\omega^{L,R}_i\right> $ represent the entangled energy eigenstates of the two regions and $\gamma$ represents the relevant length scale. Note that we have kept the superscripts $L$ and $R$ to distinguish the in and out states. This is quite a remarkable conclusion from very general considerations and must be true in any quantum theory of gravity. In the semiclassical limit, 
\begin{equation}
C_i(\omega_i, \gamma) \rightarrow e^{-\pi\omega_i/\kappa}.
\end{equation}
In this vacuum as well $\phi\left|0\right> \neq 0$. The symmetry Eq.~(\ref{genginvariance}) is not an invariance of the vacuum and only the unitarity condition is preserved. 
%%%%%%%%%%%%%%%%%%%%%%%%%%%%%%%%%%%%%%%%%%%%%%%%%%%%%%%%%%%%%%%%%%%%%%%%%%%%%%%%%%%
\subsection{Interpreting the Gauge-like, Correlation Introducing Transformation, and Connection with Final State Projection Models}
\label{complementarity}
In this subsection we will provide an interpretation of the transformation Eq.~(\ref{genginvariance}). The transformation itself has the appearance of a gauge transformation which leaves the unitarity condition $S^{\dagger}S = 1$ invariant but changes the S matrix covariantly. Gauge invariances are generally associated with redundancies in a system and this aspect was not clear in the manner that Eq.~(\ref{genginvariance}) was first derived. We will attempt to clarify this here. 

Consider the action of an S matrix on an in state to produce the corresponding out state, (recall that $g=g_1+g_2$):
\begin{equation}
\langle{\Psi_{out}(g)}| = \langle{\Psi_{in}(g)}|S(g).
\label{state}
\end{equation}
 We can introduce the generalized Hamiltonian and write the evolution of $S(g)$ in the by now familiar form:
 \begin{equation}
i{\delta S \over \delta g(x)} = \left(i{\delta S \over \delta g(x)}S^{\dagger}\right)S(g).
\label{deltaS}
\end{equation}
We may now decide to change the in state by a unitary transformation $\omega(g_1,g_2)$ which changes the correlation between the in and out states, i.e.,
\begin{equation}
\langle{\overline{\Psi}_{in}}| =  \langle{\Psi_{in}}|\omega(g_1,g_2).
\label{overline}
\end{equation}
If we want to reach the same out state as before, we must now also change the S matrix correspondingly, so that we may write,
\begin{equation}
\langle{\Psi_{out}}| = \langle{\overline{\Psi}_{in}}|S^{\prime} =  \langle{\Psi_{in}}|\omega(g_1,g_2)S^{\prime} =  \langle{\Psi_{in}}|S.
\label{relation}
\end{equation}
Clearly, $S^{\prime} = \omega^{\dagger}S$, and the S matrix itself transforms covariantly such that its unitarity is maintained provided a unitary $\omega$ may be found. In fact what Eq.~(\ref{relation}) is telling us is that $S'$ is the S matrix obtained by introducing correlations through $\omega$ into the S matrix $S$. Crucially, by construction, $S'$ is unitary if $S$ is.

We may write this for two such transformations in the following way:
\begin{align}
\langle{\Psi_{out}}| = \langle{\Psi_{in}}|\omega_1S_1 = \langle{ \Psi_{in}}|\omega_2S_2 =  \langle{\Psi_{in}}|S. \nonumber \\
\omega_1S_1 = \omega_2S_2.
\label{repeated}
\end{align}
Let us see how this redundancy gives rise to the invariance Eq.~(\ref{genginvariance}). To proceed, we first note that 
\begin{equation}
S_2 = (\omega_2^{\dagger}\omega_1)S_1 \equiv \Omega^{\dagger} S_1, 
\label{Omega}
\end{equation}
and the two evolutions are described by Eq.~(\ref{deltaS}) with $S_1$ and $S_2$ substituted for $S$.
The evolutions of $S_2(g)$ and $S_1(g)$ are related by,
\begin{align}
i{\delta S_2 \over \delta g(x)} = i{\delta \Omega^{\dagger} \over \delta g(x)}S_1(g) + \Omega^{\dagger}i{\delta S_1 \over \delta g(x)}, \nonumber \\
=  \left[i{\delta \Omega^{\dagger} \over \delta g(x)}\Omega + \Omega^{\dagger}\left(i{\delta S_1 \over \delta g(x)}S_1^{\dagger}\right)\Omega\right]S_2.
\end{align}
where we have used Eq.~(\ref{Omega}) and additionally, Eq.~(\ref{deltaS}) for $S_1$ in the second term on the right hand side of the first equation. The generalized hamiltonians are therefore related by,
\begin{eqnarray}
{\delta S_2 \over \delta g(x)}S_2^{\dagger} = \Omega^{\dagger}\left({\delta S_1 \over \delta g(x)}S_1^{\dagger}\right)\Omega + {\delta \Omega^{\dagger} \over \delta g(x)}\Omega,
\label{gaugetr}
\end{eqnarray}
which is just the transformation law Eq.~(\ref{genginvariance}) provided we identify $\Omega(g_1,g_2)$ with the horizon S matrix, $e^{-i\phi}$.

Two features of this transformation are worth emphasizing. The first is that in general, it cannot be interpreted as just a canonical transformation of the generalized hamiltonian. Indeed, the second term is the all important generalized horizon hamiltonian, or more generally the boundary hamiltonian, and for it to be nontrivial, the corresponding $\phi$ must satisfy the entangling boundary condition Eq.~(\ref{bc}). Two hamiltonians related by a canonical transformation will give a nontrivial change in the S matrix only due to boundary effects \cite{Chisholm}. We need something more than that in order to provide correlations between the in and out states: a nontrivial contribution from the second term of Eq.~(\ref{gaugetr})subject to the entangling boundary condition. In the context of the discussion at the end of section~(\ref{Modcausal}) where we introduced the substitution $\delta g(x) \rightarrow \delta(\tau-x^0)\delta \tau$, we may say that the ``time" dependance of $\phi$ and the imposition of the entangling boundary conditions are the crucial distinguishing factors. The close relation of all this to the change in the causality condition is not quantitatively seen from the above discussion and for that we must refer to section~(\ref{Modcausal}). However, in the following paragraph we discuss this in more detail. Another interesting feature is associated with the identification of the transformation implied in Eq.~(\ref{overline}) with being implemented by the boundary S matrix or in the case of black holes, the horizon S matrix. This again seems natural because the correlations implied by this transformation affect both the incoming and outgoing particles and hence they will be implemented by the horizon S matrix. Our previous analysis has indicated that this correlation may be thought of as causing a change in the shape of the horizon. Hence, it is natural to interpret the transformation, Eq.~(\ref{genginvariance}), as representing a change in the shape of the horizon. This statement however, has an important implication: black hole horizons have the property that they change with the gravitational coupling, therefore the ``boundary"  that comprises the boundary S matrix must also have this property. Obvious candidates in the context of string theory are of course the D-Branes \cite{Polchinski} and similar objects. Then the change to a new generalized hamiltonian indicated by Eq.~(\ref{genginvariance}) would be just that of the new theory including similar objects as the D-Branes.

Now, we may view the transformations of Eq.~(\ref{overline}) and Eq.(\ref{relation}) in at least two ways. One may view them, as we have been doing uptil now as a tranformation between two theories, one with and one without a horizon in which the the latter is introduced to provide the correlations between the in and out states. There is therefore an element of acausality introduced here since now the new ``in" state $\langle{\overline{\Psi}_{in}}|$ has knowledge about the out states. There is another way one may view these transformations: They could also implement the correlations between the infalling matter and the Hawking radiation in a theory with a horizon already present, to realize the final state boundary condition proposal of \cite{maldacena}. It is known \cite{preskill1,preskill2} that this proposal is an implementation of teleportation in the context of post-selected quantum mechanics and gives rise to situations with information flow to the past . Thus we see that our formulation is general enough to include the realization of the the final state boundary condition proposal of \cite{maldacena} and therefore it is not surprising that we are seeing a very similar situation with respect to acausal information flow. Indeed, the final state projection models, black hole evaporation and the horizon S matrix, and preselection are all described by the same formulation as discussed in this paper. The causality violation in each case is described by our Eq.~(\ref{geometriccausality}) and Eq.~(\ref{newgendynamics2}) where $e^{i\phi}$ is the relevant entangling unitary transformation. Only the placement of the entangling unitary operator is different for the different scenarios: Eq.~(\ref{newgenSfactor}) or more specifically, $S' = e^{i\phi} S$ best describes the positioning of the unitary for the post-selection scenario, Eq.~(\ref{thooftfactor}) gives the placement for the case of the horizon S matrix and $S' = Se^{i\phi}$ describes preselection. Our formulation gives a unified description of all of them. We will return to this discussion in the next section.

Since the effects of spacetime curvature, the entanglement across the horizon and the long range correlations are introduced through the horizon S matrix we have here an explicit realization  of the idea of ``Spacetime from Entanglement" which was also proposed from different viewpoints in \cite{Swingle,Maldacena2}. In this context, we note that 
there is a connection between the transformation, Eq.~(\ref{overline}) subject to the entangling boundary conditions and the permutation group. Indeed, consider first the finite permutation group $S_n$. It is well known that any cycle in $S_n$ is a product of transpositions. For example, a $k$-cycle may be written as,
\begin{equation}
(j_1j_2......j_k) = (j_1j_2)(j_2j_3).....(j_{k-1}j_k).
\end{equation}
Further, any permutation in $S_n$ is a product of cycles which in turn is a product of transpositions. However, there is a redundancy in the decomposition of a permutation into transpositions , unlike the cycles which are unique. Our discussion above is very suggestive that the the transformation Eq.~(\ref{overline}), with the entangling boundary conditions represents a continuous infinite permutation group. Though there is some discussion of this object in the mathematical literature \cite{cameron}, this field is still developing. How specifically the invariance, Eq.~(\ref{genginvariance}), is manifested for the corresponding transpositions and the connection of the infinite permutation groups to the structure of quantum gravity is a fascinating subject.
To the author, a positive and heartening aspect of this connection is that the permutation symmetries may be introduced outside of any discussion of specific space-time topologies. In this way, perhaps, one may argue for a close relation of the quantum gravity theory to topological field theories.  

%%%%%%%%%%%%%%%%%%%%%%%%%%%%%%%%%%%%%%%%%%%%%%%%%%%%%%%%%%%%%%%%%%%%%%%%%%%%%%%%%%%
 \subsection{Causality, Effective Field Theories and Commutation Relations}
 \label{causalityeft}
 The two terms on the RHS of Eq.~(\ref{integratedfuturedynamics}) are of a very different nature. The integral in the second term is only over the future directed light cone because the Hamiltonian $H(x,g)$ satisfies the causality requirement of Eq.~({\ref{hcausality}). Thus this part of the evolution equation respects the standard causal ordering. The first term however, has no such restriction because the horizon hamiltonian does not obey any condition like Eq.~({\ref{hcausality}). Indeed if it did, the horizon S matrix $\omega(g_1,g_2)$ would have to be of the factorized form $\omega_1(g_1)\omega_2(g_2)$ which would not satisfy the all important entangling boundary conditions Eq.~(\ref{bc}). Thus it is this term needed for the correlations between the in and out states, which however has the potential to create mischief with respect to the standard causal ordering. In the previous section we limited ourselves to the semiclassical regime where we considered only very low energy incoming modes at the horizon. Then we may assume the standard causal ordering and use Eq.~(\ref{gee}) to replace $\delta g(x)$ by the same time as in the second term and then note that these time derivatives are small. This neglect of the first term then gave the Hawking result which is thermal with no correlations between the in and out states as expected. We have thus been able to use an effective field theory description in this situation  which is valid both near the horizon and for the out states. Can we use effective field theory for determining what an observer at the horizon (infalling observer) and what the outside observer at a large distance from it (null infinity) will see, in general, not just for the low frequency incoming modes at the horizon? 
 
Before proceeding it is worthwhile to point out that quantum mechanics may allow for nonstandard causal ordering even in the absence of closed time-like curves in the geometry of spacetime \cite{Deutsch,lloyd}. In fact, a good example where information flow to the past is opened up is provided when teleportation is considered within post-selected quantum mechanics. This was discussed in \cite{preskill1,preskill2} in the context of the Horowitz-Maldacena proposal \cite{maldacena}. There it was argued that 
an overall description in terms of a consistent causal ordering could be chosen (perhaps with some fine tuning). However, at intermediate times a trace is left behind such that temporary cloning of states and polygamous entanglement could be allowed. We will argue that the same could also be expected near the horizon of a black hole even outside the explicit post-selected quantum mechanical framework. In fact, we saw in the previous section that the entangling unitary operator which enforces the correlations needed to impose the final state boundary conditions is in fact just $e^{i\phi}$. Hence even though the interpretations are different (in one case $e^{i\phi}$ is the horizon S matrix and in the other another kind of entangling unitary), the same formulation is applicable in both cases and therefore the same kind of Bogoliubov causality violation is expected. Much of the discussion of Ref.~\cite{preskill2} maybe carried over with the appropriate changes in interpretation. However, it should be emphasized that the mechanism of causality violation is very different when we view $e^{i\phi}$ as the horizon S matrix. We will elaborate on this next by outlining the nature and physical origin of causality violation in this case.

An unusual feature of the horizon S matrix which was discussed earlier (see Eq.~(\ref{phiexpansion}))and explicitly seen in the example discussed in the appendix is that $\phi$ retains knowledge of the entire history. Consider a measurement of a typical Hawking radiation made by an outside observer and let us examine the history of these modes near the horizon by extrapolating backwards in time, i.e., $t \rightarrow t-T$. The frequencies of these Hawking modes will be blue shifted by a factor $e^{T \over 4M}$. What this means is that in leaving the gravitational potential well of the black hole this same Hawking quanta's frequency was red shifted down to the smaller typical value at time $T$. Thus as the Hawking quanta are emerging from the gravitational potential well they have extremely large momenta near the horizon. This should result in a gravitational shock wave which will shift the trajectories of the infalling matter both spatially and temporally. Indeed, we may think of this as a change in shape of the horizon. This is in fact the original argument proposed in \cite{thooft1,stephens,thooft1.5} to imply correlations between the in and out states and verified in the eikonal approximation. It is this, in fact ,which also supports the modification of Bogoliubov causality that we are have found here. Note that the modification is near the horizon and hence is relevant for what we have been calling the horizon S matrix which describes the interactions between the incoming matter and the outgoing radiation. This is in agreement with what we have stated many times before: if one looks at Eq.~(\ref{integratedfuturedynamics}) it is only the first term which can allow for any change of the usual causal ordering because the integral in the second term is along the future directed light cone. This means that only for the horizon S matrix (at intermediate times), for the time dependent cases, can there be situations where the information flow can be backwards in time with the possibility of information loss only if closed time-like curves of information flow are formed. A signal of the formation of such closed causal curves would be that some additional singularities would be produced. An example where in fact this does happen is discussed in \cite{caldarelli,Milanese} in the AdS/CFT context. These singularties would spoil the hermiticity of $\phi$ and hence the unitarity of the horizon S matrix. However, if there are no effective closed time like curves of information flow, then we can provide an equivalent description of the same process away from the horizon with a standard causal ordering by ``straightening out " the bends in the flow. For this to happen, the full quantum gravity theory must therefore be either assumed  to posses symmetries or conservation laws which forbid such closed time like curves or the dynamics should be finely tuned. Once the process is straightened out, unitarity in both senses will be conserved. The horizon S matrix which plays the role of the entangling transformation will therefore now be unitary by itself. A simple example is provided by the 't Hooft calculation of the horizon S matrix outlined in section (\ref{Hawking}). There the transfer of information involves one particular quantum number, namely the radial momentum. Recall that in the eikonal approximation the radial momenta are strongly coupled but the transverse momenta are treated semiclassically. Thus momentum conservation would not allow any closed causal loops. The resulting expression for the S matrix, Eq.~(\ref{thooft-Smatrix}), is seen to be perfectly unitary. An example of appealing to fine tuning to avoid closed timelike curves of information flow is provided in \cite{preskill2}. In fact our discussion here closely parallels \cite{preskill2} where the question of noncausal propagation in the black hole evaporation problem was first addressed in the context of the Horowitz-Maldacena \cite{maldacena} proposal.  
Now the main point is that even though the overall process has a unitary S matrix in this new basis, 
traces of the noncausal propagation can still be manifested at intermediate times. These will allow for the temporary violations of the no-cloning theorem and the principle of monogamy of entanglement in a very similar manner to that discussed in \cite{preskill2}. The implications for the firewall problem \cite{Firewalls,Firewalls1} then follow: The outside observer will see a firewall like structure at the horizon but the infalling observer does not because for her the no cloning and monogamy of entanglement theorems are not valid.

One other important feature of the above discussion must be emphasized. This is that the resetting of the causal arrow must be continuously done as long as the black hole is still evaporating. Once it is completely evaporated, of course the horizon disappears. Thus if we are examining the S matrix describing the formation and {\emph{complete}} evaporation of a black hole, the usual formalism may work but to understand unitarity at any intermediate stage of black hole evaporation the formalism described in this paper appears essential. We also have uncovered a special feature that any quantum gravity theory must possess, namely,  causal ordering may be observer dependent. By this we mean that the observer, say at the horizon and the observer outside the black hole will see a different causal ordering for the same phenomenon. The special theory of relativity introduced the notion that chronological ordering is observer dependent which becomes apparent when considering speeds close to that of light. In the same sense, quantum gravity injects the role of the observer into causal ordering of phenomenon which again becomes important under extreme situations like near a black hole.

Let us next turn to the related problem of the limitations of effective field theory descriptions for the outside and the infalling observers. The approach parallels \cite{Verlinde2}, however, the emphasis here is on causality and so the implementations are different. The arguments of the previous paragraph clearly indicate that the coordinate systems of the two observers are related. Suppose that we want to have a causal effective field theory description at the horizon. In this case we will have to cut off the high frequency modes near the horizon. However, as we have seen above from the argument of the previous paragraph, this will remove modes even at moderate energies which are  important to the outgoing observer and which in fact are relevant for the leading gravitational corrections to the outgoing states. It would be the equivalent of throwing the baby out with the bath water.  If we want to give a reasonable description of the states for the outside observer, we must therefore keep the high energy modes there which will however destroy any effective field theoretical description at the horizon. The only choice that can give an effective field theory description both for the outside observer and the horizon is when we limit the {\emph{incoming modes}} at the horizon to low energy and this was precisely the choice that neglects the first term of Eq.~(\ref{integratedfuturedynamics}) and reproduces the Hawking result as discussed in section (\ref{Hawking}). 
In the ``straightened out" basis, the overall process is unitary so the out S matrix, $S_{\text out}$, and the in S matrix, $S_{\text in}$, must also be unitary by themselves. Therefore the corresponding observables must each generate a complete basis for the Hilbert spaces. The incompatibility of the descriptions inside and outside the black hole is manifested by the fact that, the corresponding observables acting on the same spacelike slice do not commute. This is because in the ``straightened" out basis the outside and inside observables act on the same system but at different times in the new causal ordering. In this way complementarity is regained. 

The question of the commutativity of local operators at space-like separations can also be viewed more formally which we explain below.
The essential problem  is that if the Hilbert space is split into factors commensurate with Eq.~(\ref{newgenSfactor}) then the in and out Hilbert spaces are not independent. From Eq.~(\ref{newgendynamics2}) and its counterpart for the in (past) states we may again try to analyse this problem of the commutativity of operators for space-like separation. For the case of  $\phi=0$, i.e., for an observer who does not see the horizon or feel its effects, we could conclude that they do in fact commute because in this case the in and out states are independent. However, when $\phi \neq 0$, there will be correlations between the in and out states then and we cannot arrive at the same conclusion. Indeed, let us consider the commutator of an essentially ``flat space" operator like the combination (see Eq.~(\ref{causality-prime})(Just for this part we revert back to the old notation of using $S$ to represent the S matrix without correlations and $S'$ the new S matrix with effects of horizon included),
\begin{equation} 
 {\delta S \over \delta g(x)}S^{\dagger}= e^{-i\phi}{\delta S' \over \delta g(x)}S'^{\dagger}e^{i\phi} - e^{-i\phi}{\delta e^{i\phi} \over \delta g(x)},
\label{invagain}
\end{equation}
It is easy to check that,
\begin{align}
{\delta H(x,g) \over \delta g(y)} = {\delta \over \delta g(y)}\left({{\delta S \over \delta g(x)}S^{\dagger}}\right) = \nonumber \\
-\left[e^{-i\phi}{\delta S' \over \delta g(x)}S'^{\dagger}e^{i\phi} - e^{-i\phi}{\delta e^{i\phi} \over \delta g(x)}\right] 
\left[e^{-i\phi}{\delta S' \over \delta g(y)}S'^{\dagger}e^{i\phi} - e^{-i\phi}{\delta e^{i\phi} \over \delta g(y)}\right] + \nonumber \\
e^{-i\phi}{\delta^2 S' \over \delta g(y) \delta g(x)}S'^{\dagger}e^{i\phi} - e^{-i\phi}{\delta^2 e^{i\phi} \over \delta g(y) \delta g(x)} + \nonumber \\
\left[e^{-i\phi}{\delta e^{i\phi} \over \delta g(x)}e^{-i\phi}{\delta S' \over \delta g(y)}S'^{\dagger}e^{i\phi}
+ (x \leftrightarrow y)\right] \nonumber \\
- \left[e^{-i\phi}{\delta e^{i\phi} \over \delta g(x)}e^{-i\phi}{\delta e^{i\phi} \over \delta g(y)}
+ (x \leftrightarrow y)\right].
\end{align}
The expression for ${\delta \over \delta g(x)}({{\delta S \over \delta g(y)}S^{\dagger}})$ can be obtained from the above by the interchange $x \leftrightarrow y$. A little algebra then allows us to conclude,
\begin{equation}
{\delta H(x,g) \over \delta g(y)} - {\delta H(y,g) \over \delta g(x)} = i\left[ H(x,g), H(y,g)\right] 
\label{compatibility}
\end{equation}
Recall that $H(x,g)$ is essentially a flat space operator living in the usual Fock space.
 Consequently with Lorentz invariance, for space like separations, the commutator on the right hand side vanishes and microcausality is preserved. Note that this just a statement of the compatibility condition for  the left hand side of Eq.~(\ref{invagain}) (see Eq.~(\ref{microcausality})) which satisfies the usual causality condition. However, all other operators not occurring in this combination, for example, ${\delta S' \over \delta g(x)}S'^{\dagger}$ or $({\delta e^{i\phi} \over \delta g(x)})e^{-i\phi}$,  considered separately do not live in a Fock space and need not commute at space like separations. In these cases the in and out states are correlated in the presence of the horizon and are therefore not independent. Thus even though one may derive a similar relation as Eq.~(\ref{compatibility}) with $H'(x,g)={\delta S' \over \delta g(x)}S'^{\dagger}$ replacing $H(x,g)$, no such compatibility condition can be used here to conclude that $H'(x,g)$ and $H'(y,g)$ will commute for spacelike separations.

 %%%%%%%%%%%%%%%%%%%%%%%%%%%%%%%%%%%%%%%%%%%%%%%%%%%%%%%%%%%%%%%%%%%%%%%%%%%%%%%%%%%
\subsection{Connection with the Faddeev-Kulish Asymptotic States Theory}
\label{FKconnection}

Let us now see how Eq.~(\ref{integratedfuturedynamics}) can be applied to the study of infrared divergences in QED in flat space. The reason for this is to reproduce a known result \cite{FK} obtained historically using a different approach. We will use field theoretical considerations to find $\phi$ in order to construct the infrared S matrix in QED. In this way we hope to convince the reader of the correctness of our approach, in particular, the validity of Eq.~(\ref{integratedfuturedynamics}). Thus we will use for $H$ in the second term on the RHS of Eq.~(\ref{integratedfuturedynamics}), the known QED hamiltonian relevant for infrared physics and construct $\phi$ and hence the asymptotic states of this theory. As mentioned earlier, using a completely different  approach Faddeev and Kulish \cite{FK} showed how to construct the same infrared finite S matrix for QED. In their approach the asymptotic scattering states are governed not by the free time evolution operator but by a different one , $U_{\text{asym}}$ which takes into account the non-vanishing of the QED hamiltonian at large times. The usual Dyson S matrix is calculated between these coherent states in which every particles is accompanied by a cloud of soft photons. In this section we will show how Eq.~(\ref{integratedfuturedynamics}) may be used to obtain these results. In this case perturbation theory can be used in the second term on the RHS of Eq.~(\ref{integratedfuturedynamics}) and the evolution equation may be explicitly solved in the infrared limit. This not only provides a check for our formalism but also teaches lessons about how to approach the evolution equation in the more difficult case of the black hole. 

Referring back to the discussion at the end of previous section, we see that in the second integral of Eq.~(\ref{integratedfuturedynamics}), we may replace $\delta g(x) = f^{\prime}(x^0-\tau) \delta \tau$. Thus,
\begin{equation}
i\delta S(g) = \int i\frac{\delta e^{i\phi}}{\delta g(x)}e^{-i\phi} S(g) \delta g(x) d^4x + \int e^{i\phi}H(x,g)e^{-i\phi}S(g) f^{\prime}(x^0-\tau) \delta \tau d^4x.
\end{equation}
Then we divide by $\delta \tau$ and in the integrand of the first integral, and define $i\frac{\delta e^{i\phi}}{\delta g(x)}e^{-i\phi}\delta g(x) = \delta \Gamma(x^0, {\mathbf{x}}, g)$. The reason we are able to say something about this integral is because it has support only along the plane $x^0 = \tau$. Thus we may make the replacement,
\begin{equation}
i\frac{\delta e^{i\phi}}{\delta g(x)}e^{-i\phi} \delta g(x) \rightarrow \delta \Gamma(x^0, {\mathbf{x}}, g)\delta(x^0-\tau).
\end{equation}
Then, defining, $\Psi(\tau,g) = \int d^3x \Gamma(\tau, {\mathbf{x}}, g) = -\phi(\tau,g) + ...$, we can rewrite Eq.~(\ref{integratedfuturedynamics}) in the desired form,
\begin{equation}
i{\partial S(\tau) \over \partial \tau}= {\partial \Psi(\tau,g) \over \partial \tau} S(\tau) + \int e^{i\phi}H(x)e^{-i\phi}d^3x S(\tau) .
\end{equation}
We are interested in the solution of this equation at large times where only the infrared sensitive terms are retained.
Denoting the effective Hamiltonian by $H_{eff}$, we have, 
\begin{equation}
H_{\text{eff}}(\tau) =  {\partial \Psi(\tau) \over \partial \tau} + \int e^{i\phi}H(x)e^{-i\phi}d^3x ,
\label{H-effective}
\end{equation}
and the S matrix is given by:
\begin{equation}
S = T e^{\displaystyle -i\left(\int_{-\infty}^{+\infty}H_{\text{eff}}(\tau) d\tau \right)}.
\end{equation}
The first term in the effective Hamiltonian contributes only at the boundary, at large times and plays the role of an asymptotic operator. This is exactly what we had demanded when the operator $\phi$ was introduced.

We next discuss the nature of the asymptotic states in this theory and point out that already here the structure of the Hilbert space is unconventional even though it is well understood.

First we must determine the asymptotic form of $\phi$ and then that of the asymptotic time evolution operator. The dominant term in the interaction Hamiltonian in the limit of very large times for the QED Hamiltonian is straightforward to determine \cite{FK}:
\begin{align}
H_{asym} = -\frac{e}{(2\pi)^{3/2}}\int \frac{d^3p d^3k}{\sqrt{2k^0}}\frac{p^{\mu}}{p^0}\left[e^{\displaystyle -i\frac{p\cdot k}{p^0}t}\rho^{\dagger}(p)a_{\mu}(k) + e^{\displaystyle i\frac{p\cdot k}{p^0}t}\rho(p)a^{\dagger}_{\mu}(k) \right] \nonumber \\
\rho(p) = \sum_{\sigma}\left(b^{\dagger}(p,\sigma)b(p,\sigma) - d^{\dagger}(p,\sigma)d(p,\sigma)\right).
\end{align}
In the above, $a_{\mu}(k)$ is the photon annihilation operator and the only nonvanishing contribution at large times comes from the soft photon momentum region when $k_{\mu} \rightarrow 0$. Hence the above expression is valid in the small $k$ and large $|t|$ regions and the corresponding limits have been used in writing $H_{asym}$ in the above form. 
Next we must choose a $\phi$ such that in the second term on the RHS of  Eq.~(\ref{H-effective}) all the infrared sensitive terms are removed. For this purpose, consider the expansion of this term,
\begin{equation}
e^{i\phi} H e^{-i\phi} = H + i [\phi, H] + \frac{i^2}{2!}[\phi,[\phi,H]] + ...
\label{expansion}
\end{equation}
We now perform a perturbative expansion in $e$, and note that this is accomplished with the condition,
\begin{equation}
i[\phi,H_0] = - H_{asym},
\end{equation}
where, $H_0$ is the free Hamiltonian. This identifies,
\begin{equation}
\phi(t) = - \int^t dt' H_{asym}(t').
\end{equation}
With this choice the order $e$  term in Eq.~(\ref{expansion}) removes the trilinear coupling in the QED Hamiltonian such that $e^{i\phi} H e^{-i\phi}$ does not contain any infrared divergences. All of these are therefore contained in the first term of Eq.~(\ref{H-effective}). In this way we get, in agreement with \cite{FK} that the asymptotic time evolution operator is,
\begin{equation}
U_{\text{asym}}(t) = T~ e^{\displaystyle -i\left(\int^t (H_0 + H_{asym}(t')) dt' \right)}.
\end{equation}
Using the Baker- Campbell-Hausdorf formula this  can be written in the form,
\begin{equation}
e^{ [-iH_0t]}e^{[i\Phi(t)]}e^{[i\phi(t)]}.
\end{equation}
In the above, $\Phi(t)$ is the coulomb phase factor, which represents the logarithmic distortion of the plane wave component of the asymptotic charged field. In scattering processes it is cancelled by contributions from the imaginary part of diagrams with virtual photons \cite{Weinberg}. The boundary S matrix  $e^{i\phi}$ constructs the space of the asymptotic states of the theory. It is responsible for the creation of a soft photon cloud attached to a charged particle. In this sense the boundary S matrix $e^{i\phi}$ again acts like a clothing or dressing operator 
as in the previous section. Specifically for $t \rightarrow \infty$ where only the soft photon momentum region contributes, 
\begin{equation}
\phi(t) =  -i\frac{e}{(2\pi)^{3/2}}\int \frac{d^3p d^3k}{\sqrt{2k^0}}\frac{p^{\mu}}{p\cdot k}\left[e^{\displaystyle -i\frac{p\cdot k}{p^0}t}\rho^{\dagger}(p)a_{\mu}(k) - e^{\displaystyle i\frac{p\cdot k}{p^0}t}\rho(p)a^{\dagger}_{\mu}(k)\right]. 
\end{equation}
When written in coordinate space this represents a nonlocal interaction. However since we are considering the infrared region only this is not surprising and neither is the causality violation implied by this choice of $\phi$. The space of asymptotic states can be essentially built from the perturbative vacuum by the action of  $e^{i\phi}$ and has been studied in \cite{FK} (see also \cite{Ware} for the case of perturbative gravity). One has to introduce a non-Fock Hilbert space by means of the coherent states. An important feature of these is that the set of coherent states is overcomplete. This is what is to be expected because as we have pointed out earlier, the in, and out states and the states on the boundary are not all independent. 

The asymptotic evolution operator constructs asymptotic fields from the free fields through a unitary transformation. For example, for the free photon field $A_{\mu}(x)$ we have,  $U_{\text{asym}} A^{\mu}U^{\dagger}_{\text{asym}} = A^{\mu}_{\text{asym}}$ .
To see some of the properties of asymptotic fields one may look at the non-zero commutator of 
$A^{\mu}_{\text{asym}}$ with the electron creation operator, $b^{\dagger}(p)$. Some algebra gives,
\begin{equation}
[A^{\mu}_{\text{asym}}(x),b^{\dagger}({\bf q})] = \frac{e}{4\pi}\frac{q^{\mu}}{\sqrt{(q\cdot x)^2-q^2x^2}}b^{\dagger}({\bf q}).
\label{lienard}
\end{equation}
One sees here the emergence of the classical Lienard-Wiechert potential on the right hand side.  Thus with respect to the asymptotic operator each charged particle is created with its electromagnetic field. The case of perturbative gravity is discussed using another approach in \cite{Ware}. There what appears on the right hand side of the corresponding commutator is the gravitational field of a particle moving with momentum $q^{\mu}$ as given by the Aichelburg- Sexl metric. These results were obtained in the semiclassical approximation. Nevertheless, they may have some bearing on how the the Maldacena-Susskind proposal \cite{Maldacena2} of a wormhole connecting the inside and outside of a black hole can be obtained in the full quantum gravity theory.
%%%%%%%%%%%%%%%%%%%%%%%%%%%%%%%%%%%%%%%%%%%%%%%%%%%%%%%%%%%%%%%%%%%%%%%%%%%%%%%%%%%

%%%%%%%%%%%%%%%%%%%%%%%%%%%%%%%%%%%%%%%%%%%%%%%%%%%%%%%%%%%%%%%%%%%%%%%%%%%%%%%%%%%
\section{Conclusions}
\label{Conclusion}
We have developed a S matrix theory of gravity and applied it, in particular, to study aspects of the information loss problem in black hole evaporation. The effects of spacetime curvature and the horizon are incorporated through the introduction of the boundary or the horizon S matrix and no particular metric or lagrangian structure is assumed. This horizon S matrix may be considered to be a generalization of the Bogoliubov transformation and plays a dual role: It generates the entanglement between the inside and outside of a black hole which drives the Hawking information loss argument and is also responsible for the S matrix unitarity preserving long range correlations between the in and out states, through gravitational and other interactions. In this way we suggest another approach to the problem of how spacetime  may be thought to arise from entanglement. 

If we want to examine the S matrix describing the formation and {\emph{complete}} evaporation of a black hole then the usual formulation may work but to understand unitarity at any intermediate stage of black hole evaporation the formalism described in this paper appears to be essential. We have constructed a procedure to introduce long range correlations between two causally complementary regions and in this way conserve information flow. Our framework is broad enough to describe both, the dynamics at the horizon of a black hole and also, how to implement the correlations needed in final state projection models. In the perturbative context it reproduces the Faddeev-Kulish theory of asymptotic dynamics needed to construct  an infrared finite S matrix. An important ingredient for providing the correlations through a unitary transformation is the choice of the entangling boundary conditions which we have proposed. Significantly, we have seen that in order to introduce the correlations necessary for unitarization, the Bogoliubov causality requirement must necessarily be changed. Using this new form of causality we find an evolution equation satisfied by the S matrix. In the black hole case, under some reasonable assumptions,  the full S matrix may be split into a product of three factors: $S_{\text{out}}S_{\text{horizon}}S_{\text{in}}$ consistent with 't Hooft's original suggestion. Of these, we find that it is the horizon S matrix that is most important. In this sense we are seeing holography at work. At the technical level we have found it convenient to use the Bogoliubov S matrix approach in which the evolution of the S matrix is studied as a function of a propagating disturbance without introducing time, nor is a foliation of spacetime necessary. The causality condition is changed in the process of introducing the horizon S matrix which is the nonlocal unitary transformation generating the correlations between spacetime regions which are causal complements. This unitary transformation is, in turn, introduced through a symmetry of the unitarity condition of the S matrix. The choice of the entangling boundary conditions  for the horizon S matrix is an essential ingredient. We would like to emphasize that we have tried to see how much information we can extract from our approach without detailed assumptions about the nature of the quantum gravity theory. In section 4 we have looked at specific applications. We have used the evolution equation to study black hole radiance and reproduced Hawking's result when limiting ourselves to low frequency incoming modes at the horizon. We have also argued how to go beyond this but using the eikonal approximation for gravity to introduce the gravitational corrections responsible for the long range correlations. Based again on the nature of the evolution equation and the entangling boundary conditions, we have suggested a form for the vacuum state at the horizon in the full quantum theory of gravity.  The effectiveness of field theoretical descriptions is also discussed. As a test of our evolution equation we have turned to the problem of infrared divergences in QED and found the implications of our approach for the asymptotic states of the theory. In this way our results reproduce the Faddeev-Kulish theory \cite{FK}. In an appendix we solve the evolution equation for a simplied case and show how information may be transferred from the in to the out states by a phase factor which stores knowledge of the entire history.

The violation of standard causality is very similar in two apparently different contexts: first, when the correlations are introduced through the horizon S matrix for the case of black hole evaporation and secondly, for the analysis of final state projection models. We have argued that as long as there are no closed timelike curves of information flow one could find an equivalent description where the standard causal ordering is recovered in the context of the S matrix description outside the horizon. At intermediate stages, however, this violation of Bogoliubov causality could allow for cloning of states and polygamy of entanglement and in this way circumvent the firewall problem. This of course places important constraints  on the dynamics of the quantum theory of gravity. One such constraint is that a hermitian $\phi$ which is the generator of the horizon S matrix does not have singularities and there are no obstructions to its construction. We have argued that if we assume that due to symmetry or fine tuning considerations, the full quantum gravity theory forbids closed timelike curves of information flow, then the three individual S matrices in the decomposition $S_{\text{out}}S_{\text{horizon}}S_{\text{in}}$ are by themselves unitary. Complementarity is then seen to follow. 

S matrix theory was popular in the 1960s in the context of strong interactions. Eventually the emphasis was changed which led to the development of string theory which currently is the best accepted approach to quantum gravity. In this paper, we have attempted to directly apply S matrix theory to the problem of quantum gravity. Some general results may be extracted without detailed assumptions about the dynamics of quantum gravity. However, many questions cannot be answered without more specifics about the interactions at the planck scale. An intriguing outcome is the possibility to describe the emergence of spacetime as arising due to entanglement. The relation of this approach to string theory is also an interesting problem for the future. The general formalism developed in this paper potentially can be applied to a diverse variety of problems, for example, it may be used to analyze decoherence in an S matrix approach.\footnote{The Referee is thanked for suggesting this possibility.} In this situation the role of the horizon S matrix will be taken over by the environment S matrix. Work in this direction is in progress.

As this paper was being readied for publication, the preprints in Ref.~\cite{Raju1,Raju2} appeared. In these seemingly elegant and thorough papers a resolution of the black hole information paradox is proposed which is tied very much with the AdS/CFT context and gives the impression that the bulk description is inadequate for this. To examine the role of the bulk more closely, it would be very interesting to make a connection of these papers with the proposal presented here.

%%%%%%%%%%%%%%%%%%%%%%%%%%%%%%%%%%%%%%%%%%%%%%%%%%%%%%%%%%%%%%%%%%%%%%%%%%%%%%%%%%%
\section*{Acknowledgements}
I would like to thank Gerard 't Hooft for patiently explaining his S matrix approach to black hole dynamics over a couple of lengthy e-mail exchanges. I am also grateful to Ashoke Sen for a quick look at the manuscript and for his comments. Finally, I sincerely thank the anonymous Referee for encouraging remarks and for suggestions which have improved not only the presentation but also the physics content.
%\end{acknowledgements}
%%%%%%%%%%%%%%%%%%%%%%%%%%%%%%%%%%%%%%%%%%%%%%%%%%%%%%%%%%%%%%%%%%%%%%%%%%%%%%%%%%%
\newpage
\appendix
\section{Explicit Structure of the Full S matrix For the Abelian Case}
\label{abeliancase}
For  Eq.~(\ref{newSfactor}), the associative property gives,
\begin{equation}
S(g_3+(g_1+g_2)) = S((g_3+g_1)+g_2),
\end{equation}
which in turn implies for the functionals $\phi$,
\begin{equation}
\phi(g_3,g_1) + \phi(g_3+g_1,g_2) = \phi(g_1,g_2) + \phi(g_3,g_1+g_2).
\label{associative}
\end{equation}
Let us introduce the total derivative,
\begin{equation}
{d \over dg} = \int d^4x {\delta \over \delta g(x)},
\label{totalderivative}
\end{equation}
and a new function $\psi(\tau,\sigma) = {d\phi(\sigma, \tau) \over d\sigma}$.
Differentiating Eq.~(\ref{associative}) with respect to $g_3$ at $g_3=0$ gives,
\begin{equation}
\psi(0,g_1+g_2) = \psi(0,g_2) + \psi(g_1,g_2).
\end{equation}
Now, let us define another function,
\begin{equation}
A(\tau) = \int_0^{\tau}\psi(0,\sigma)d\sigma. 
\label{lambda}
\end{equation}
Then it follows that,
\begin{align}
A(g_1+g_2) - A(g_2) - A(g_1) = \nonumber \\
\int_{g_2}^{g_1+g_2}\psi(0,\sigma)d\sigma - \int_0^{g_1}\psi(0,\sigma)d\sigma = \nonumber \\
\int_0^{g_1}[\psi(0,\sigma+g_2) - \psi(0,\sigma)]d\sigma = \int_0^{g_1}\psi(\sigma,g_2)d\sigma \nonumber \\
= \int_0^{g_1}{d \phi(\sigma,g_2) \over d\sigma}d\sigma = \phi(g_1,g_2) - \phi(0,g_2).
\end{align}
Using Eq.~(\ref{bc}), we conclude,
\begin{equation}
A(g_1+g_2) - A(g_1) - A(g_2) = \phi(g_1+g_2).
\end{equation}
Thus, if we define a primed S-matrix through the relation:
\begin{equation}
S(g) = e^{iA(g)}S^{\prime}(g),
\end{equation}
Then Eq.~(\ref{newSfactor}) can be written in terms of the $S^{\prime}$ matrix in the same form as Eq.~(\ref{Sfactor}), i.e., 
\begin{equation}
S^{\prime}(g_1+g_2) = S^{\prime}(g_2)S^{\prime}(g_1).
\end{equation}
One may think of the relation between the primed S-matrix and the unprimed as being realized by the following correspondence between the corresponding out states,
\begin{equation}
|\beta>_{out} \rightarrow e^{iA_{out}(g)}|\beta>_{out}.
\label{outstates}
\end{equation}
In any case, the in and out states will differ by a coupling constant dependent phase factor of $e^{iA(g)}$. Thus if the causality condition is changed to Eq.~(\ref{newcausality}) then the only effect, for the abelian case, is to change the out states by a multiplication by the phase factor $e^{iA(g)}$ which makes them coupling constant dependent in the sense that $A(g)$depends on the entire profile or history of the disturbance in the ``in" region. This is in contrast to the standard case when the out states are in one to one correspondence with the states of a free particle. 

Two other aspects of our conclusion are novel and at variance with current ideas. First, we restate that the kind of correlations between the future and past light cones suggested by Eq.~(\ref{newcausality}) lead to the result shown in Eq.~(\ref{outstates}). The picture thus suggested is what we would expect from phenomena like quantum teleportation and Eqs.~(\ref{newcausality}) and ~(\ref{outstates}) represent a novel way to implement this idea. Secondly, consider the definition of the the phase factor $A(g)$ given in Eq.~(\ref{lambda}). It has the unusual feature that it depends on the whole history of the physical system. Again, this is in contradistinction to the current dogma that the quantum mechanical wavefunction(al) encodes {\emph{all}} the facets left from past history which are responsible for future evolution. Both of the above mentioned conclusions have a commonality with the EPR paradox.
 Thus, in this simple model, we can see clearly how information is carried to the out states from the in states. The entire history is ``stored" in $A(g)$, as is seen from  Eq.~(\ref{lambda}) which involves an integral over the disturbance in the ``in" region.
We conclude this discussion of the abelian case by mentioning another aspect of 
Eq.~(\ref{outstates}). Recall that Eq.~(\ref{ginvariance}) represents an invariance of the unitarity relation. Eq.~(\ref{outstates}) gives the tranformation of the states under this invariance. Again, this transformation law has the appearance of an abelian gauge transformation on the state vectors.  

%%%%%%%%%%%%%%%%%%%%%%%%%%%%%%%%%%%%%%%%%%%%%%%%%%%%%%%%%%%%%%%%%%%%%%%%%%%%%%%%%%%
%%%%%%%%%%%%%%%%%%%%%%%%%%%%%%%%%%%%%%%%%%%%%%%%%%%%%%%%%%%%%%%%%%%%%%%%%%%%%%%%%%%

\newpage

\end{document}